\documentclass[12pt]{article}
\usepackage{fancyhdr}

\usepackage{mathrsfs}
\usepackage[T1]{fontenc}
\usepackage{mathpazo}
\usepackage{setspace}
\usepackage{amsfonts}
\usepackage{amssymb}
\usepackage{amsmath}
\usepackage{epsfig}
\usepackage{latexsym}
\usepackage{color}
\usepackage{graphicx}
\usepackage{nicefrac}
\usepackage[latin1]{inputenc}
\usepackage{slashed}
\usepackage{multirow}
\usepackage{comment}
\usepackage{soul}
\usepackage{hyperref}
\usepackage[nosort]{cite}
\usepackage{datetime}

\usepackage{braket}

\usepackage{cancel}

\usepackage[titletoc]{appendix}

\def\hybrid{\topmargin -20pt    \oddsidemargin 0pt
        \headheight 0pt \headsep 0pt
        \textwidth 6.25in       
        \textheight 9.25in       
        \marginparwidth .875in
        \parskip 5pt plus 1pt   \jot = 1.5ex}

\hybrid

\def\baselinestretch{1.2}

\catcode`\@=11

\def\marginnote#1{}
\newcount\hour
\newcount\minute
\newtoks\amorpm
\hour=\time\divide\hour by60
\minute=\time{\multiply\hour by60 \global\advance\minute by-\hour}
\edef\standardtime{{\ifnum\hour<12 \global\amorpm={am}%
        \else\global\amorpm={pm}\advance\hour by-12 \fi
        \ifnum\hour=0 \hour=12 \fi
        \number\hour:\ifnum\minute<10 0\fi\number\minute\the\amorpm}}
\edef\militarytime{\number\hour:\ifnum\minute<10 0\fi\number\minute}

\def\draftlabel#1{{\@bsphack\if@filesw {\let\thepage\relax
   \xdef\@gtempa{\write\@auxout{\string
      \newlabel{#1}{{\@currentlabel}{\thepage}}}}}\@gtempa
   \if@nobreak \ifvmode\nobreak\fi\fi\fi\@esphack}
        \gdef\@eqnlabel{#1}}
\def\@eqnlabel{}
\def\@vacuum{}
\def\draftmarginnote#1{\marginpar{\raggedright\scriptsize\tt#1}}

\def\draft{\oddsidemargin -.5truein
        \def\@oddfoot{\sl preliminary draft \hfil
        \rm\thepage\hfil\sl\today\quad\militarytime}
        \let\@evenfoot\@oddfoot \overfullrule 3pt
        \let\label=\draftlabel
        \let\marginnote=\draftmarginnote
   \def\@eqnnum{(\theequation)\rlap{\kern\marginparsep\tt\@eqnlabel}%
\global\let\@eqnlabel\@vacuum}  }

\def\preprint{\twocolumn\sloppy\flushbottom\parindent 2em
        \leftmargini 2em\leftmarginv .5em\leftmarginvi .5em
        \oddsidemargin -.5in    \evensidemargin -.5in
        \columnsep .4in \footheight 0pt
        \textwidth 10.in        \topmargin  -.4in
        \headheight 12pt \topskip .4in
        \textheight 6.9in \footskip 0pt
        \def\@oddhead{\thepage\hfil\addtocounter{page}{1}\thepage}
        \let\@evenhead\@oddhead \def\@oddfoot{} \def\@evenfoot{} }

\def\numberbysection{\@addtoreset{equation}{section}
        \def\theequation{\thesection.\arabic{equation}}}

\def\underline#1{\relax\ifmmode\@@underline#1\else
        $\@@underline{\hbox{#1}}$\relax\fi}

\def\titlepage{\@restonecolfalse\if@twocolumn\@restonecoltrue\onecolumn
     \else \newpage \fi \thispagestyle{empty}\c@page\z@
        \def\thefootnote{\fnsymbol{footnote}} }

\def\endtitlepage{\if@restonecol\twocolumn \else \newpage \fi
        \def\thefootnote{\arabic{footnote}}
        \setcounter{footnote}{0}}  

\catcode`@=12
\relax

\def\figcap{\section*{Figure Captions\markboth
        {FIGURECAPTIONS}{FIGURECAPTIONS}}\list
        {Figure \arabic{enumi}:\hfill}{\settowidth\labelwidth{Figure
999:}
        \leftmargin\labelwidth
        \advance\leftmargin\labelsep\usecounter{enumi}}}
 \relax
\def\tablecap{\section*{Table Captions\markboth
        {TABLECAPTIONS}{TABLECAPTIONS}}\list
        {Table \arabic{enumi}:\hfill}{\settowidth\labelwidth{Table
999:}
        \leftmargin\labelwidth
        \advance\leftmargin\labelsep\usecounter{enumi}}}
 \relax
\def\reflist{\section*{References\markboth
        {REFLIST}{REFLIST}}\list
        {[\arabic{enumi}]\hfill}{\settowidth\labelwidth{[999]}
        \leftmargin\labelwidth
        \advance\leftmargin\labelsep\usecounter{enumi}}}
 \relax
\makeatletter
\newcounter{pubctr}
\def\publist{\@ifnextchar[{\@publist}{\@@publist}}
\def\@publist[#1]{\list
        {[\arabic{pubctr}]\hfill}{\settowidth\labelwidth{[999]}
        \leftmargin\labelwidth
        \advance\leftmargin\labelsep
        \@nmbrlisttrue\def\@listctr{pubctr}
        \setcounter{pubctr}{#1}\addtocounter{pubctr}{-1}}}
\def\@@publist{\list
        {[\arabic{pubctr}]\hfill}{\settowidth\labelwidth{[999]}
        \leftmargin\labelwidth
        \advance\leftmargin\labelsep
        \@nmbrlisttrue\def\@listctr{pubctr}}}
 \relax
\makeatother
\newskip\humongous \humongous=0pt plus 1000pt minus 1000pt

\newif\ifdtup

\relax

\def\be{\begin{equation}}
\def\ee{\end{equation}}
\def\ba{\begin{eqnarray}}
\def\ea{\end{eqnarray}}

\def\del{\partial}

\def\r{\rho}
\def\a{\alpha}

\def\b{\beta}

\def\g{\gamma}

\def\d{\delta}

\def\th{\theta}

\def\m{\mu}

\def\om{\omega}

\def\l{\lambda}

\def\s{\sigma}

\def\cL{{\cal L}}

\def\no{\noindent}

\def\qq{\qquad}

\def\IR{\relax{\rm I\kern-.18em R}}

\def \z { {\bar z} }

\def \ha {{1\over 2}}

\def \ov {\over}

\def\diag{{\rm diag}}

\def\IR{\relax{\rm I\kern-.18em R}}
\def\IL{\relax{\rm I\kern-.18em L}}

\def\inv{^{\raise.15ex\hbox{${\scriptscriptstyle -}$}\kern-.05em 1}}

\def\cL{{\cal L}}

\def\Tr{{\rm Tr}}

\begin{document}

\renewcommand{\theequation}{\thesection.\arabic{equation}}
\csname @addtoreset\endcsname{equation}{section}

\newcommand{\beq}{\begin{equation}}
\newcommand{\eeq}[1]{\label{#1}\end{equation}}
\newcommand{\ber}{\begin{equation}}
\newcommand{\eer}[1]{\label{#1}\end{equation}}
\newcommand{\eqn}[1]{(\ref{#1})}
\begin{titlepage}
\begin{center}


${}$
\vskip .2 in

\vskip .4cm

{\large\bf
A free field perspective of  $\lambda$-deformed coset CFT's 
 }

\vskip 0.5in

{\bf George Georgiou},\hskip .2cm {\bf Konstantinos Sfetsos}\hskip .17cm   and \hskip .1cm  {\bf Konstantinos Siampos}
\vskip 0.2in

 {\em
Department of Nuclear and Particle Physics,\\
Faculty of Physics, National and Kapodistrian University of Athens,\\
Athens 15784, Greece\\
}

\vskip 0.14in

{\footnotesize \texttt {george.georgiou, ksfetsos, konstantinos.siampos@phys.uoa.gr}}

\vskip .5in
\end{center}

\centerline{\bf Abstract}

\no
We continue our study of $\lambda$-deformed $\sigma$-models by setting up a $\nicefrac{1}{k}$ perturbative expansion around the free field point for cosets, in particular for the  $\lambda$-deformed $SU(2)/U(1)$ coset CFT. We construct an interacting field theory in which all deformation effects are manifestly encoded in the interaction vertices. Using this we reproduce the known $\beta$-function and the anomalous dimension of the composite operator perturbing away from the conformal point. We introduce the $\lambda$-dressed parafermions which have an essential Wilson-like phase in  their expressions. 
Subsequently, we compute their anomalous dimension, as well as their four-point functions, as exact functions of the deformation and to leading order in the $k$ expansion. Correlation functions with an odd number of these parafermions  vanish as in the conformal case.

\vskip .4in
\noindent
\end{titlepage}
\vfill
\eject

\newpage

\tableofcontents

\noindent

\def\baselinestretch{1.2}
\baselineskip 20 pt
\noindent


\setcounter{equation}{0}

\section{Introduction }

A class of integrable two-dimensional field theories having an explicit action realization, 
were systematically constructed in recent years \cite{Sfetsos:2013wia,Hollowood:2014rla,Hollowood:2014qma,Georgiou:2016urf,Georgiou:2017jfi,Sfetsos:2017sep,Sfetsos:2015nya,Georgiou:2018hpd,Georgiou:2018gpe,Driezen:2019ykp}.  They typically represent the effective action corresponding to
 the deformation of one or more WZW model current algebra theories at levels $k_i$ by current bilinears.  
They are called $\lambda$-deformed $\sigma$-models due to the preferred letter used to denote the deformation parameters.
In this context, a research avenue is the systematic study of various aspects of the corresponding two-dimensional quantum field  theories. Roughly speaking, these fall into two general categories. 
In the first, belong studies concerning directly the coupling constants in these theories. 
In particular, the computation of their running under the renormalization group flow (RG) ($\beta$-functions) has been exhaustively 
studied \cite{Kutasov:1989dt,Gerganov:2000mt,Itsios:2014lca,Sfetsos:2014jfa,Appadu:2015nfa,Sfetsos:2017sep,Sagkrioti:2018rwg,Georgiou:2018hpd,Sagkrioti:2018abh}. In addition, the geometrical aspects of the space of couplings in these theories have been elucidated \cite{Georgiou:2015nka,Georgiou:2019jcf} and the $C$-function capturing the number of the degrees of freedom along the RG flow has been evaluated \cite{Georgiou:2018vbb,Sagkrioti:2018abh}. 
The second category consists of works aiming at discovering how the operators of the CFT respond to the deformation. 
That includes the computation of the anomalous dimension they acquire  \cite{Georgiou:2015nka,Georgiou:2016iom,Georgiou:2016zyo,Georgiou:2017oly,Georgiou:2017aei} as well as their 
dressing induced by the deformation of the original CFT \cite{Georgiou:2019jcf}.
The above works utilized a combination of CFT and gravitational techniques together with symmetry arguments in the couplings 
space of these models. The results that were obtained are valid to all orders in the deformation parameters and to leading order 
at the level $k$. 

Recently a new purely field theoretic approach to the study of $\lambda$-deformed theories was initiated in \cite{Georgiou:2019aon}. This method was applied to the isotropic case having a single coupling $\lambda$. The resulting theory is a theory of free fields having certain interaction terms with all the coupling constants depending on $\lambda$,
in a specific manner.  Similarly, the dressed operators of the theory, elementary as well composite ones, can be expressed, building also on results of \cite{Georgiou:2019jcf},  in terms of the free fields with specific couplings.
In this approach all computations are organized around a free field theory and not around the conformal 
point. The advantage is that all information about the deformation parameter $\lambda$  is encoded in the coupling constants appearing in the action
and in the various coefficients in the expressions of the operators which are given in terms of the free fields.  
This approach delivered results for the $\beta$-function, correlation functions and anomalous dimensions in complete agreement with the previous methods \cite{Georgiou:2015nka,Georgiou:2019jcf}, however with much less effort. 

Most of the studies in this direction were done for the $\lambda$-deformed WZW current algebra 
CFTs. However, $\lambda$-deformations can be constructed for coset CFTs which have an action 
realization based on gauged WZW models. In particular, $\lambda$-deformed models have been constructed based on
the $SU(2)/U(1)$ coset CFT in \cite{Sfetsos:2013wia}, for more general symmetric spaces in \cite{Hollowood:2014rla} 
and the $\text{AdS}_5 \times S^5$ superstring in \cite{Hollowood:2014qma}.
In the present paper we will construct and utilize 
the field theory based on free fields, analog of the construction for the group case of  \cite{Georgiou:2019jcf} for the $\lambda$-deformed  $SU(2)/U(1)$ coset CFT. With our approach we will be able to compute the anomalous dimension of 
parafermionic fields which are seemingly very hard to do by means of other methods. In particular, it is much more difficult
to apply conformal perturbation theory when the underlying theory is a coset CFT instead of a current algebra CFT. 
Essentially, this is due to the fact that  in the former case parafermions are involved instead of currents and they have
more complicated  correlation functions \cite{Fateev:1985mm}.

The plan of the paper is as follows: In section \ref{sec2}, the perturbative expansion of the action around the free field
point will be performed. In this way we will construct an interacting field theory by keeping  terms up to sixth order in the fields which suffices for our purposes. In the process we will freely use various field redefinitions in order to simplify the final form
of the action. In section \ref{sec3}, we will use  this action to compute the $\beta$-function of the $\sigma$-model using 
standard heat-kernel techniques.  In section \ref{sec4}, we will introduce the $\lambda$-dressed parafermions which contain an important Wilson-like phase in their expressions. Next, we compute the anomalous dimension of the composite operator  which is bilinear in the
parafermions  and which drives the model away from
the conformal point. We also compute the anomalous dimension of a single parafermion. In this case, the presence of the
non-trivial Wilson-like phase plays the important r\^ole. In addition, we compute all four-point functions of parafermions. 
Finally, in section \ref{sec5} we will present our conclusions and future directions of this work. Last but not least, two appendices follow.
In appendix \ref{appA} we calculate the anomalous dimension of variations of the composite operator deforming the CFT while in appendix \ref{appB} a list of integrals and 
the regularization scheme we employ is considered.

\section{Constructing the interacting theory}
\label{sec2}

In this section the interacting field theory corresponding to the $\lambda$-deformed coset CFT for $SU(2)/U(1)$ will be constructed.
Our approach follows in spirit that of \cite{Georgiou:2019aon}. In that work the interacting theory, based on free fields,
for the isotropic single $\lambda$-deformed $\sigma$-model deformation of the WZW model CFT  was constructed.

\subsection{Expansion around the free point}

Even though we are interested in the coset case, it is convenient to start  with the general $\lambda$-deformed $\sigma$-model action  for the group case given by \cite{Sfetsos:2013wia}
\be
\label{sjksdsjd}
 S_{k,\l}(g)=S_\text{WZW,k}(g)+\frac{k}{\pi}\int \text{d}^2\s\, R^a_+(\l^{-1}-D^T)_{ab}^{-1}L^b_-\,,
\ee
where
\be
S_{\text{WZW},k}(g)=-\frac{k}{2\pi}\int\text{d}^2\s\,\text{Tr}(g^{-1}\del_+ gg^{-1}\del_-g)+\frac{k}{12\pi}\int\text{Tr}(g^{-1}\text{d}g)^3\,,
\ee
is the WZW action for a group element $g$  of a semi-simple Lie group $G$ at level $k$ and
\be
R^a_+=-i\,\text{Tr}(t_a\del_+ gg^{-1})\,,\quad L^a_-=-i\,\text{Tr}(t_ag^{-1}\del_- g)\,,\quad D_{ab}=\text{Tr}(t_agt_bg^{-1})\, .
\ee
The $t_a$'s are representation matrices satisfying
\be
[t_a,t_b]=if_{abc}t_c\,,\quad \text{Tr}(t_at_b)=\d_{ab}\,,\quad a=1,2,\dots,\text{dim G}\,,
\ee
where the $f_{abc}$'s are the algebra structure constants which are taken to be real.
The coupling matrix $\l_{ab}$ parametrizes the deviation from the conformal point at which the currents $R_+$ and $L_-$ are
chirally, respectively anti-chirally, conserved.

\no
The $\lambda$-deformed action for the $SU(2)/U(1)$ coset CFT can be obtained from the above action specialized to the $SU(2)$ case with 
$t_a=\s_a/\sqrt{2}$, where the $\s_a$'s are the Pauli matrices, and $\l_{ab}=\diag(\l,\l,\l_3)$ where $\l_3$ corresponds to the $U(1)$ subgroup of $SU(2)$ via a limiting procedure \cite{Sfetsos:2013wia}. We review this  by first parameterizing the group element as
\be
\label{lllg}
g=\text{e}^{i(\phi+\phi_1)\s_3/2}\text{e}^{i(\pi/2-\th)\s_2}\text{e}^{i(\phi-\phi_1)\s_3/2}\,,
\ee
where the range of values of the Euler angles are
\be
 \th\in[0,\nicefrac{\pi}{2}]\,,\quad \phi\in[0,2\pi]\,,\quad \phi_1\in[0,2\pi]\,.
\ee
Inserting the above into \eqref{sjksdsjd} and taking the limit $\l_3\to1$ we obtain that
\be
\label{actioncoset}
\begin{split}
 S_{k,\l}(g)= &{k\ov \pi}  \int \text{d}^2\s\,  \bigg({1-\l\ov 1+\l}(\del_+\th \del_-\th + \tan^2\th \del_+\phi \del_-\phi)
\\
&+{4\l\ov 1-\l^2}(\cos\phi \del_+\th -\sin \phi \tan\th \del_+\phi) (\cos\phi \del_-\th -\sin \phi \tan\th \del_-\phi)\bigg)\, .
\end{split}
\ee
We note that the coordinate $\phi_1$ has decoupled at the level of the action and the remaining two fields are $\th$ and $\phi$. However, this angle will play a very important r\^ole, as we shall see in the course of the paper. Its presence will be instrumental in determining the form of the $\lambda$-dressed parafermions and, as a consequence, of their anomalous dimensions.

\no
The action \eqref{actioncoset}  is invariant under the following two symmetries \cite{Kutasov:1989aw,Itsios:2014lca}
\be
\label{sksldkdqq}
\begin{split}
{\rm I}: &\qq \l\to\l^{-1}\,,\quad k\to-k\,,\\
{\rm II}: &\qq \l\to-\l\,,\quad \phi\to\phi+\frac{\pi}{2}\,.
\end{split}
\ee
In what follows we will see that these symmetries will be manifest in the expressions for the physical quantities of this theory.

\no
Next we proceed with our construction by zooming around $\th=0$ by first setting
\be
\label{zooomm}
\th= {\r\ov \sqrt{2 k}}\
\ee
and defining two new fields as
\be
\label{polar}
y_1 = \sqrt{1+\l\ov 1-\l}\, \r \cos\phi\ , \qq y_2 = \sqrt{1-\l\ov 1+\l}\, \r \sin\phi\ .
\ee
Then, in the large $k$ expansion the action becomes
\be
\label{acctt}
\begin{split}
 S_{k,\l} = {1\ov 2\pi} &\int \text{d}^2\s\, \Big(\del_+y_1\del_-y_1 + \del_+y_2\del_-y_2
+ {g_{11}\ov k}\, y_2^2 \,\del_+y_1\del_-y_1
\\
& + {g_{22}\ov k}\, y_1^2 \,\del_+y_2\del_-y_2
+   {g_{12}\ov k}\, y_1 y_2 \, (\del_+y_1\del_-y_2+ \del_+y_2\del_-y_1)
\\
&
+{y_2^2\ov k^2}\, (h_{11} y_2^2 + \tilde h_{11} y_1^2)\, \del_+y_1\del_-y_1
+{y_1^2\ov k^2}\, (h_{22} y_1^2 + \tilde h_{22} y_2^2)\, \del_+y_2\del_-y_2
\\
& +{y_1 y_2\ov k^2}\, (h_{12} y_1^2 + \tilde h_{12} y_2^2)\, (\del_+y_1\del_-y_2+ \del_+y_2\del_-y_1)
\Big)
+\cdots\ ,
\end{split}
\ee
where we have kept terms up to quartic order in the fields.
Note that, this is an expansion in the number of fields. Retaining 
appropriate powers of $\nicefrac1k$, is just for book keeping as we could rescale the fields by a factor of $\sqrt{k}$  and
have $k$ as an overall coefficient in the action. 
The various couplings are given by
\be
\label{coopl}
\begin{split}
& g_{11}=  {1\ov 3}{1+ \l\ov 1-\l}\ ,\qq g_{22}=  {1\ov 3}{1- \l\ov 1+\l}\ ,\qq
g_{12}= - {1\ov 3}{1+ \l^2\ov 1-\l^2}\ ,
\\
& h_{11}={17 (1+\l)^2\ov 180 (1-\l)^2}\ ,\qq   \tilde h_{11}= {17+ 14 \l + 17 \l^2\ov 180 (1+\l)^2}\ ,
\\
& h_{22}={17 (1-\l)^2\ov 180 (1+\l)^2}\ ,\qq   \tilde h_{22}= {17- 14 \l + 17 \l^2\ov 180 (1-\l)^2}\ ,
\\
& h_{12}= -{17- 10 \l + 17 \l^2\ov 180 (1+\l)^2}\ ,\qq
\tilde h_{12}=-{17 +  10 \l + 17 \l^2\ov 180 (1-\l)^2}\ .
\end{split}
\ee
Note that \eqref{acctt} with the above couplings is invariant under the transformations
\be
\label{fsyyklkm}
\begin{split}
{\rm I}: &\qq  \l\to {1\ov \l}\ ,\qq\ \ k\to -k\ ,\\
{\rm II}: & \qq \l\to -\l\ ,\qq (y_1,y_2) \to (-y_2,y_1)\ .
\end{split}
\ee
which of course correspond to \eqref{sksldkdqq} when the above zoom-in limit is taken.

\no
Next we perform the following field redefinitions
\be
\label{field.redefinition}
y_1=  \Big(1 +{a_1\ov k}  x_2^2 + {b_1\ov k^2} x_2^4 + {c_1\ov k^2} x_1^2 x_2^2 \Big)x_1\ ,
\quad y_2=  \Big(1+{a_2\ov k}  x_1^2 + {b_2\ov k^2} x_1^4 + {c_2\ov k^2} x_1^2 x_2^2 \Big)x_2\ .
\ee
Choosing the coefficients as
\be
\label{coopl1}
\begin{split}
& a_1 = -{g_{11}\ov 2}\ ,\qq b_1= {3 g_{11}^2\ov 8} - {h_{11}\ov 2}\ ,
\qq c_1 = {g_{11}g_{22}\ov 6} +{g_{12}g_{22}\ov 3} -{g_{22}^2\ov 6} -{\tilde h_{11}\ov 6}\ ,
\\
&
 a_2 = -{g_{22}\ov 2}\ ,\qq b_2= {3 g_{22}^2\ov 8} - {h_{22}\ov 2}\ ,
\qq c_2 = {g_{22}g_{11}\ov 6} +{g_{12}g_{11}\ov 3} -{g_{11}^2\ov 6} -{\tilde h_{22}\ov 6}\ ,
\end{split}
\ee
we obtain the simpler action
\be
\label{acctt10}
\begin{split}
 S_{k,\l} = {1\ov 2\pi} &\int \text{d}^2\s\, \Big(\del_+x_1\del_-x_1 + \del_+x_2\del_-x_2
+   {\hat g_{12}\ov k}\, x_1 x_2 \, (\del_+x_1\del_-x_2+ \del_+x_2\del_-x_1)
\\
& +{x_1 x_2\ov k^2}\, (\hat h_{12} x_1^2 + \hat {\tilde h}_{12} x_2^2)\, (\del_+x_1\del_-x_2+ \del_+x_2\del_-x_1)
\Big)
+\cdots\ ,
\end{split}
\ee
where the new couplings are denoted by a hat and are given by
\be
\label{coopl2}
\begin{split}
& \hat g_{12} = g_{12} - g_{11}-g_{22} = - {1+\l^2\ov 1-\l^2} \ ,
\\
&
\hat h_{12} = {1\ov 3} \big(g_{11}g_{22} - g_{12}g_{22} + 2 g_{22}^2  + 3 h_{12} - 6 h_{22} - \tilde h_{11}\big)
= -{1\ov 6}\, \bigg({1-\l\ov 1+\l}\bigg)^{\!2}\ ,
\\
&
\hat {\tilde h}_{12} = {1\ov 3} \big(g_{22}g_{11} - g_{12}g_{11} + 2 g_{11}^2  + 3 \tilde h_{12} - 6 h_{11} - \tilde h_{22}\big)
=  -{1\ov 6}\, \bigg({1+\l\ov 1-\l}\bigg)^{\!2}\ .
\end{split}
\ee
Obviously, the action \eqref{acctt10} with the above couplings is invariant under \eqref{fsyyklkm},
where for the symmetry II the $y_a$'s should be replaced accordingly by the $x_a$'s.

\subsection{Computational QFT conventions}

We would like to set up a perturbative expansion around the free theory and perform quantum computations. Passing to the
Euclidean regime we have the following basic propagators which are consistent with our normalizations
\be
\langle x_a(z_1,\z_1) x_b(z_2,\z_2) \rangle= -\d_{ab}\ln |z_{12}|^2\ ,\qq a=1,2\ ,
\ee
where $z_{12}= z_1-z_2$.
Note that the above propagator implies that
\be
\label{noott}
\langle \del x_a(z_1)\bar\del x_b(\z_2) \rangle=\pi\, \d_{ab}\d^{(2)}(z_{12})\,,
\ee
inducing a coupling of the holomorphic and anti-holomorphic sectors which will be very important  in the calculations 
that follow, in particular in subsubsections \ref{4pt1} \& \ref{4pt5}.

\no
We will see later that it will be most convenient to define two complex conjugate bosons as
\be
x_\pm = x_1 \pm i x_2\ .
\ee
In this complex basis, the only non-vanishing two-point function is
\be
\langle x_+(z_1,\z_1) x_-(z_2,\z_2)\rangle = -2 \ln |z_{12}|^2\ .
\ee
Finally, for the free theory the holomorphic energy--momentum tensor is given by
\be
T= -\ha \Big((\del x_1)^2 + (\del x_2)^2\Big) = -\ha \del x_+ \del x_- \ ,
\ee
with a similar expression for the anti-holomorphic one.

\section{The $\beta$-function}
\label{sec3}

In this section we use the field theory action \eqref{acctt10} in order to compute the $\beta$-function for the deformation
parameter $\lambda$. The analogous computation for the group case was performed in \cite{Georgiou:2019aon}.
We will find precisely the result of \cite{Itsios:2014lca} obtained by gravitational methods.

\no
In order to obtain the $\beta$-function for $\lambda$ we will employ the background field heat kernel method, so that as an initial step we need the equations of motion for the fields $x_1$ and $x_2$  
derived from the action \eqref{acctt10}.
We obtain, up to ${\cal O}(\nicefrac{1}{k^2})$, that
\be
\label{eqs1}
\begin{split}
&
\del_+ \del_- x_1 + {x_1 x_2\ov k} \bigg(\hat g_{12}+{\hat h_{12} x_1^2 + \hat {\tilde h}_{12} x_2^2\ov k}\bigg)\,
 \del_+ \del_- x_2
 \\
 &\qq\qq\qq\qq\qq + {x_1\ov k} \bigg(\hat g_{12}+{\hat h_{12} x_1^2 +3 \hat {\tilde h}_{12} x_2^2\ov k}\bigg)\,
\del_+ x_2 \del_-x_2 =0\ ,
\\
&
\del_+ \del_- x_2 + {x_1 x_2\ov k} \bigg(\hat g_{12}+{\hat h_{12} x_1^2 + \hat {\tilde h}_{12} x_2^2\ov k}\bigg)\,
 \del_+ \del_- x_1
 \\
& \qq\qq\qq\qq\qq + {x_2\ov k} \bigg(\hat g_{12}+{3\hat h_{12} x_1^2 + \hat {\tilde h}_{12} x_2^2\ov k}\bigg)\,
\del_+ x_1 \del_-x_1 =0\ .
\end{split}
\ee
For large $k$ we may simplify them by solving for $\del_+\del_- x_a$, $a=1,2$. Keeping terms up to ${\cal O}(\nicefrac{1}{k^2})$, we find that
\be
\label{eqs2}
\begin{split}
&\del_+ \del_- x_1 + {\hat g_{12}\ov k} x_1 \del_+ x_2 \del_- x_2
\\
&\qq\qq\qq
+{x_1\ov k^2} \Big(\big(\hat h_{12} x_1^2 + 3 \hat{\tilde h}_{12} x_2^2\big)\,
\del_+ x_2 \del_- x_2 - \hat g_{12}^2 x_2^2\, \del_+ x_1 \del_- x_1\Big) =0\ ,
\\
&\del_+ \del_- x_2 + {\hat g_{12}\ov k} x_2 \del_+ x_1 \del_- x_1
\\
&\qq\qq\qq
+{x_2\ov k^2} \Big(\big(\hat{\tilde h}_{12} x_2^2 + 3 \hat h_{12} x_1^2\big)\,
\del_+ x_1 \del_- x_1 - \hat g_{12}^2 x_1^2\, \del_+ x_2 \del_- x_2\Big) =0\ .
\end{split}
\ee
From \eqref{eqs2} one can derive the equations which the fluctuations $\d x_a$ of the coordinates $x_a$ obey. The fluctuations 
will be taken around a classical solution which we will still denote by $x_a$.
They will be casted in the form
\be
\label{dhatt}
\hat D_{ab}\,\d x_b = 0\ ,
\ee
where the operator $\hat D$ is a certain second order in the worldsheet derivatives that will also depend on the classical solution around which we expand.
We will present its explicit expression after performing the Euclidean analytic continuation and passing to momentum space. In the conventions of \cite{Georgiou:2018hpd}, we replace $(\del_+,\del_-) $ by $\nicefrac12 (\bar p,p)\equiv (p_+,p_-)$. Then after dividing by $p_+p_-$ we obtain for $\hat D$ the result
\be
\hat D_{ab} = \d_{ab} + {1\ov k} \Big(\hat F_2 +  {\hat F'}_2+\frac1k\hat F_4\Big)_{ab}\,,
\ee
where the matrices are given by
\be
\begin{split}
 &\hat F_2 ={\hat g_{12}\ov p_+ p_-} \left(
                              \begin{array}{cc}
                                \del_+ x_2 \del_- x_2 & 0 \\
                                0 & \del_+ x_1\del_-x_1\\
                              \end{array}
                            \right)\ ,
\\
& \hat F_2' =\hat g_{12} \left(
                              \begin{array}{cc}
                               0&  {x_1\del_+ x_2\ov p_+}+  {x_1\del_- x_2\ov p_-} \\
                                 {x_2\del_+ x_1\ov p_+}+  {x_2\del_- x_1\ov p_-} & 0\\
                              \end{array}
                            \right)\ ,
\end{split}
\ee
and
\be
\label{F4terms}
\begin{split}
& \hat F_4 ={1\ov p_+p_-} \left(
                              \begin{array}{cc}
                           (\hat F_4)_{11} &    (\hat F_4)_{12}
                               \\
                               (\hat F_4)_{21} &   (\hat F_4)_{22}\\
                              \end{array}
                            \right)\ ,
\\
& (\hat F_4)_{11} = 3\big(\hat h_{12} x_1^2 +  \hat{\tilde h}_{12} x_2^2\big)\,
\del_+ x_2 \del_- x_2 - \hat g_{12}^2 x_2^2\, \del_+ x_1 \del_- x_1\ ,
\\
&  (\hat F_4)_{22}  =3\big(\hat {\tilde h}_{12} x_2^2 +  \hat h_{12} x_1^2\big)\,
\del_+ x_1 \del_- x_1 - \hat g_{12}^2 x_1^2\, \del_+ x_2 \del_- x_2\,.
\end{split}
\ee
Note that we have not provided the expressions for $(\hat F_4)_{12}$ and $(\hat F_4)_{21}$, since their form 
will be irrelevant for the discussion that follows.
Integrating out the fluctuations, gives the effective Lagrangian of our model which reads
\be
\label{lefff}
-\cL_{\rm eff} = \cL^{(0)}_{k,\l} + \int^\m {\text{d}^2 p\ov (2\pi)^2} \ln (\det \hat D)^{-1/2}\ ,
\ee
where $\cL^{(0)}_{k,\l}$  is the Lagrangian \eqref{acctt10} on the classical solution.
This integral is logarithmically divergent with respect to the UV mass scale
$\m$. The logarithmic term is isolated by performing the large momentum expansion of the integrand
and keeping terms proportional to $\nicefrac{1}{|p|^2}$, where $|p|^2=p\bar p$.
Next we use the fact that
\be
\ln (\det \hat D)  = {1\ov k} \Tr\hat F_2 + {1\ov k^2}\left( \Tr\hat F_4  - {1\ov 2}  \Tr\hat F_2^{`2}\right) + \dots \ ,
\ee
where we have included only terms potentially contributing to the logarithmically divergent term.
Then calculating the traces individually one obtains
\be
\begin{split}
&
\Tr \hat F_2= {\hat g_{12}\ov p_+p_-}\, \big(\del_+ x_1\del_-x_1 +   \del_+ x_2\del_-x_2\big)\ ,
\\
&
\Tr \hat F_4= {1\ov p_+p_-}\Big[ \big((3\hat {\tilde h}_{12}-\hat g_{12}^2) x_2^2 + 3 \hat h_{12} x_1^2\big)\,
\del_+ x_1 \del_- x_1
\\
&\qq\qq\qq
 + \big((3 \hat h_{12}-\hat g_{12}^2) x_1^2 + 3 \hat {\tilde h}_{12} x_2^2\big)\,
\del_+ x_2 \del_- x_2\Big]\ ,
\\
&
\Tr\hat F_2^{`2} =  2  \hat g_{12}^2 {x_1 x_2\ov p_+p_-} \big(\del_+ x_1\del_-x_2 +   \del_+ x_2\del_-x_1\big) \ ,
\end{split}
\ee
where again we have not included terms which will vanish upon the angular integration that follows.
Using polar coordinates, i.e. $p=re^{i\om}$, $\bar p=re^{-i\om}$, in which
the integration measure is  $\text{d}^2 p = r\text{d}r\text{d}\om$, we evaluate the effective action from \eqref{lefff} to be
(we return back to the Lorentzian regime)
\be
\label{hk2ui1}
\begin{split}
& S_{\rm eff}= {1\ov 2\pi} \int \text{d}^2\s\, \bigg[\Big(1-{\hat g_{12}\ov k}\ln \m^2\Big)
\big(\del_+ x_1 \del_-x_1 + \del_+ x_2 \del_-x_2\big)
\\
& \qq\qq +  {\hat g_{12}\ov k}\, x_1 x_2 \, \Big(1+{\hat g_{12}\ov k}\ln \m^2\Big) \big(\del_+x_1\del_-x_2+ \del_+x_2\del_-x_1\big)
\\
& \qq\qq -{\ln \m^2 \ov k^2}\Big[ \big( 3 \hat h_{12} x_1^2+(3\hat {\tilde h}_{12}-\hat g_{12}^2) x_2^2 \big)\,
\del_+ x_1 \del_- x_1
\\
&\qq\qq\qq\quad
 + \big(3 \hat {\tilde h}_{12} x_2^2+ (3 \hat h_{12}-\hat g_{12}^2) x_1^2 \big)\,
\del_+ x_2 \del_- x_2\Big]
\bigg]+\cdots\ .
\end{split}
\ee
The wavefunction renormalization and field redefinition
\be
\begin{split}
 & x_1 = \Big(1+ {\hat g_{12}\ov 2 k}\ln \m^2\Big)\bigg(1+ 
 {\hat h_{12} \hat x_1^2 + (3 \hat {\tilde h}_{12}-\hat g_{12}^2) \hat x_2^2\ov 2 k^2} \ln \m^2\bigg) \hat x_1\,, \\
 & x_2 = \Big(1+ {\hat g_{12}\ov 2 k}\ln \m^2\Big)\bigg(1+ {\hat {\tilde h}_{12} \hat x_2^2 
 + (3 \hat h_{12}-\hat g_{12}^2) \hat x_1^2\ov 2 k^2} \ln \m^2\bigg) \hat x_2\,,
\end{split}
\ee
puts the kinetic term into a canonical form and \eqref{hk2ui1} becomes
\be
\begin{split}
\label{hk2ui2}
& S_{\rm eff}= {1\ov 2\pi} \int \text{d}^2\s\, \bigg[\del_+ \hat x_1 \del_- \hat x_1
+ \del_+ \hat x_2 \del_- \hat x_2
\\
&\qquad
+{1\ov k} \Big(\hat g_{12}+{ \hat g_{12}^2  + 3 \hat h_{12} + 3 \hat{\tilde h}_{12}\ov k}\ln \m^2\Big)
 \hat x_1 \hat x_2 \,
\big(\del_+ \hat x_1\del_-\hat x_2+ \del_+\hat x_2\del_-\hat x_1\big)\bigg]+\cdots\ .
\end{split}
\ee
Demanding now that the action \eqref{hk2ui1} is $\m$-independent, i.e. $\frac{\text{d}\cL_{\rm eff}}{\text{d}{\ln \m^2}}=0$, and keeping in mind that we should keep for consistency the leading term in the $\nicefrac{1}{k}$ expansion
we obtain
\be
\label{systrg0}
{\text{d}  {\hat g}_{12} \ov \text{d}\ln \m^2}= - {1\ov k} \big(\hat g_{12}^2 + 3 \hat h_{12} + 3  \hat {\tilde h}_{12} \big) \ .
\ee
Using the specific expression for the couplings \eqref{coopl2} we finally get that
\be
\label{systrg}
\b^\l = {\text{d} \l\ov \text{d}\ln \m^2} =  -{\l\ov  k}\ .
\ee
We have, thus, reproduced the result of \cite{Itsios:2014lca}.

\section{Parafermion correlators}
\label{sec4}

The objects naturally arising in the $SU(2)/U(1)$  coset CFT are the parafermions \cite{Fateev:1985mm}. These will be denoted by
$\Psi_\pm$ and $\bar\Psi_\pm$ for the holomorphic and anti-holomorphic sectors, respectively.
In this section, we will calculate correlation functions, namely two and four-point functions, for  the $\lambda$-deformed versions of the
CFT parafermions. At the CFT point all correlation functions with an odd number of parafermions vanish.
We will show that the same statement is true for the deformed versions of the correlators, as well. The first question we address  in subsection \ref{para}
is how the deformation dresses the chirally conserved  $SU(2)/U(1)$ CFT parafermions which have a classical
form in terms of the fields $\th$ and $\phi$  \cite{Bardacki:1990wj} .
Equipped with their correct form  we proceed to
calculate in subsection \ref{bilin} the anomalous dimension of the bilinear in the parafermions operator that perturbs the
theory  away from the free point. Our result is in complete agreement with the one obtained
earlier in \cite{Georgiou:2019nbz} by the use of methods involving the geometry in the coupling space.
In the next two subsections,  we calculate the anomalous dimension of the single $\lambda$-dressed
parafermion and all four-point correlation functions involving parafermionic fields.
We will develop methods to deal with Wilson-line factors present in the expressions for the parafermion fields.
All our results will respect the symmetries \eqref{sksldkdqq} of the action.

\subsection{$\lambda$-dressed parafermionic fields}
\label{para}

 In the $\sigma$-model \eqref{sjksdsjd}, derived in \cite{Sfetsos:2013wia} throughout a certain gauging procedure,
 the classical equations of motion for the  gauge fields assume the following form
\be
\label{gaugefields}
A_+=i(\l^{-T}-D)^{-1}R_+\,,\quad A_-=-i(\l^{-T}-D^T)^{-1}L_-\, .
\ee
As discussed in \cite{Georgiou:2016iom,Georgiou:2019jcf}, these $\lambda$-dependent fields are the counterparts of the chiral and anti-chiral currents of the conformal point to which they reduce, up to overall scales, after taking  the limit $\l\to 0$.
Hence they provide the correct form of the operator generalizing the chiral and anti-chiral currents in the presence of $\lambda$.
The above form was essential for obtaining the correct anomalous dimensions and correlation functions
of currents in the free field based approach of \cite{Georgiou:2019aon} and this will be the case here, as well.

The aim of this work is to apply the approach of \cite{Georgiou:2019aon} to the case where the deformed theory is not a deformation  of a group based CFT but of a coset CFT, and more specifically of the  $SU(2)/U(1)$ coset CFT.
In this case  the natural chiral objects of the CFT are not currents but parafermions.  As it happens for the currents in the group case, the parafermions are dressed when the deformation parameter $\lambda$ is turned on.
The way the parafermions are dressed  should be derived from the same limiting
procedure we used to obtain the action.
In order to proceed,  recall that we have parametrized the group element of $SU(2)$ as in \eqref{lllg}  and that we have chosen the deformation matrix to take the form $\l_{ab}=\diag(\l,\l,\l_3)$, where $\l_3$ corresponds to the Abelian subgroup $U(1)$ in $SU(2)$. In order to obtain the $\sigma$-model
action \eqref{actioncoset} one should set the parameter $\l_3=1$.
In this procedure the Euler angle $\phi_1$ appearing in the group element
\eqref{lllg} drops out of the $\sigma$-model
which instead of being three-dimensional becomes two-dimensional \eqref{actioncoset}.
We should follow the same limiting procedure for the gauge fields \eqref{gaugefields} as well.
In doing so, we firstly and most importantly realize that the gauge fields $A^a_\pm$, $a=1,2,3$ retain an explicit
dependence on the angle $\phi_1$ which therefore at that level does not decouple.
In fact, we will see that this is a desired feature for the
classical description of the parafermions even at the CFT point. In particular, we find that, after passing to the Euclidean regime, the gauge fields projected, via the above limiting procedure, to the coset take the form
\be
\label{gaugetopara}
\begin{split}
& A_+^1 = \frac{\l}{\sqrt{2}} \Big(\Psi_+ -\Psi_-\Big)\ ,\qq A_+^2 =- i\frac{\l}{\sqrt{2}} \Big(\Psi_+ +\Psi_-\Big)\,,\\
&
 A_-^1 = -\frac{\l}{\sqrt{2}} \Big(\bar \Psi_+ -\bar \Psi_-\Big)\ ,\qq A_-^2 = i\frac{\l}{\sqrt{2}} \Big(\bar \Psi_+ +\bar \Psi_-\Big)\  ,
\end{split}
\ee
where the $\Psi$'s, as we will see, will be the $\lambda$-deformed parafermions.
They are given in terms of the angles parametrizing the group element $g$, as follows after the analytic continuation to the Euclidean regime 
\be
\label{skdsjjds}
\begin{split}
&\Psi_+=\frac{1}{1-\l^2}\left((\text{e}^{-i\phi}+
\l\text{e}^{i\phi})\del\th-i(\text{e}^{-i\phi}-\l\text{e}^{i\phi})\tan\th\del\phi\right)\text{e}^{-i\phi_1}\,,\\
&\Psi_-=\frac{1}{1-\l^2}\left((\text{e}^{i\phi}+
\l\text{e}^{-i\phi})\del\th+i(\text{e}^{i\phi}-\l\text{e}^{-i\phi})\tan\th\del\phi\right)\text{e}^{i\phi_1}\,,\\
&\bar\Psi_+=\frac{1}{1-\l^2}\left((\text{e}^{i\phi}+
\l\text{e}^{-i\phi})\bar\del\th+i(\text{e}^{i\phi}-\l\text{e}^{-i\phi})\tan\th\bar\del\phi\right)\text{e}^{-i\phi_1}\,,\\
&\bar\Psi_-=\frac{1}{1-\l^2}\left((\text{e}^{-i\phi}+
\l\text{e}^{i\phi})\bar\del\th-i(\text{e}^{-i\phi}-\l\text{e}^{i\phi})\tan\th\bar\del\phi\right)\text{e}^{i\phi_1}\,.
\end{split}
\ee
The components of the gauge fields along the subgroup $U(1)$ turn out to be
\be
\label{skslshaoq}
A_+^3=\frac{i}{\sqrt{2}}\left(\del\phi_1+J\right)\,,\qquad
A_-^3=\frac{i}{\sqrt{2}}\left(\bar\del\phi_1-\bar J\right)\,,
\ee
where the angle $\phi_1$ has become  imaginary after the analytic continuation to the Euclidean regime and $J$ and $\bar J$ are given by
\be
\label{JJc}
\begin{split}
&J=\frac{ \left(\left(1-2\l\cos\phi+\l^2\right)\tan\th\,\del\phi\
-2 \l\sin2\phi\,\del\th\right)\tan\th}{1-\l^2}\,,\\
&\bar J=\frac{\left(\left(1-2\l\cos\phi+\l^2\right)\tan\th\,\bar\del\phi-2 \l\sin2\phi\,\bar\del\th\right)\tan\th}{1-\l^2}\, .
\end{split}
\ee
Under the non-perturbative symmetries \eqref{sksldkdqq} the $\Psi$'s transform as
\be
\begin{split}
 {\rm I}: & \qq \Psi_\pm\to-\l\text{e}^{\mp2i\phi_1}\Psi_\mp\ ,\qquad \bar\Psi_\pm\to-\l\text{e}^{\mp2i\phi_1}\bar\Psi_\mp\,,\\
{\rm II}: &\qq \Psi_\pm\to\mp i\,\Psi_\pm\ ,\qquad \bar\Psi_\pm\to\pm i\,\bar\Psi_\pm\,.
\end{split}
\ee
In addition, the gauge fields satisfy the following equations of motion \cite{Hollowood:2014rla}
\be
\label{eomgauge}
\begin{split}
&\del A^{g/h}_-=-[A^{g/h}_-,A^h_+]\,,\quad \bar\del A^{g/h}_+=-[A^{g/h}_+,A^h_-]\,,
\\
&\del A^h_- - \bar\del A^h_+=\l^{-1}[A^{g/h}_+,A^{g/h}_-]\,,
\\
&A^h_\pm=A_\pm^3t_3\,,\quad A^{g/h}_\pm=A_\pm^\a t_\a\,,\quad \a=1,2\,.
\end{split}
\ee

\no
In order to elucidate the expressions for the $\Psi$'s we firstly consider  the conformal limit $\l=0$.
Then, these become the standard classical parafermions of the coset $SU(2)/U(1)$ CFT that are given by \cite{Bardacki:1990wj}
\be
\label{claparaf}
\Psi_\pm=\left(\del\th\mp i\tan\th\del\phi\right)\text{e}^{\mp i(\phi+\phi_1)}\,,\quad
\bar\Psi_\pm=\left(\bar\del\th\pm i\tan\th\bar\del\phi\right)\text{e}^{\pm i(\phi-\phi_1)}\ .
\ee
Using \eqref{gaugetopara} and \eqref{eomgauge}, we find that the gauge fixing conditions $A_\pm^3=0$ imply that the parafermions $\Psi_\pm$ and  
$\bar\Psi_\pm$  are on-shell chirally and anti-chirally conserved, respectively\footnote{The second of
 \eqref{eomgauge} seems singular in the $\l=0$ limit. However, it is also satisfied since $A_\pm^{g/h}\sim\l$,  so that $\l^{-1}[A^{g/h}_+,A^{g/h}_-]\sim\l$, as $\l\to 0$.}
\be
\bar\del\Psi_\pm=0\ ,\qquad \del\bar\Psi_\pm=0\,.
\ee
In addition, the constraints $A_\pm^3=0$ enforces $\phi_1$ to satisfy the following equations
\be
\label{ssfjdsshdrur}
\begin{split}
&\del\phi_1=-J_0\ ,\qquad \bar\del\phi_1=\bar J_0\ ,
\\
&J_0=\tan^2\th\,\del\phi\ ,\qquad \bar J_0=\tan^2\th\,\bar\del\phi\ .
\end{split}
\ee
Note that $J_0$ and $\bar J_0$ can be obtained from \eqref{JJc} after setting $\l=0$.
Moreover, using \eqref{ssfjdsshdrur}
one can show that $\phi_1$ satisfies on-shell the compatibility condition
\be
\label{isometry}
\left(\del\bar\del-\bar\del\del\right)\phi_1=\del\bar J_0+\bar\del J_0=0\,.
\ee
The above conservation law results from the $U(1)$ isometry of the $SU(2)/U(1)$ coset CFT.
Given \eqref{isometry}, one can solve
\eqref{ssfjdsshdrur}
to express the angle $\phi_1$ as a line integral\footnote{In a closed curve it would identically vanish
due to Stokes theorem and \eqref{isometry}
$$
\frac{i}{2}\oint_C\left(-\text{d}z\, J_0+\text{d}\bar z\,\bar J_0\right)=\int_S\text{d}^2z\left(\del\bar J_0+\bar\del J_0\right)=0\,,\quad C=\del S\,.
$$}
\be
\label{phase}
\phi_1(z,z_0)=\phi_1(z)-\phi_1(z_0)=\int_C\left(\del\phi_1\text{d}z\,+\bar\del\phi_1\text{d}\bar z\right)=\int_C\left(-J_0\text{d}z +\bar J_0\text{d}\bar z\right)\,,\quad
\ee
where $C$ is a curve connecting the arbitrary base point $(z_0,\bar z_0)$ with the end point $(z,\bar z)$.
Note that, given the above, the phase $\phi_1$ is independent from the choice of $C$ but depends only the base and end points.
However, correlation functions should not depend on the arbitrary base point. In the following sections, we will see that this is indeed the case.

However, in the case of non-zero deformation $\l\neq0$ the conditions  $A_\pm^3=0$ can no longer be imposed, as can be seen from the second equation in \eqref{eomgauge}. Nevertheless,
one can impose an alternative gauge fixing condition, namely
\be
A^3_+=-\frac{i}{\sqrt{2}}\,F\,,\quad A^3_-=\frac{i}{\sqrt{2}}\,\bar F\,,
\ee
where $F$ is at the moment an arbitrary function which will be later specified and $\bar F$ is its complex conjugate. 
Equivalently, using  \eqref{skslshaoq}, we find
\be
\label{skslshaoq1}
\begin{split}
\del\phi_1=-J_F\,,\quad
\bar\del\phi_1=\bar J_F\,,
\end{split}
\ee
with $J_F=J+F$. The equations \eqref{skslshaoq1} uniquely determines $\phi_1$ on-shell provided that the following consistency condition is satisfied
\be
\label{sksjsqqdks}
\left(\del\bar\del-\bar\del\del\right)\phi_1=\del\bar J_F+\bar\del J_F=0\,.
\ee
Equivalently, the last equation implies that $F$  and $\bar F$ should satisfy the relation
\begin{equation}
\del\bar F+\bar\del F=\frac{2\l}{1-\l^2}\left((\del\th\bar\del\th-\del\phi\bar\del\phi\tan^2\th)\sin2\phi
+(\del\th\bar\del\phi+\del\phi\bar\del\th)\tan\theta\cos2\phi\right)\,.
\end{equation}
Similarly to \eqref{phase}, we can solve \eqref{skslshaoq1} and \eqref{sksjsqqdks} and express  $\phi_1$ as a line integral through
\be
\label{phase1}
\phi_1(z,z_0)=\phi_1(z)-\phi_1(z_0)=\int_C\left(-J_F\text{d}z+\bar J_F\text{d}\bar z\right)\,.
\ee
In what follows, we shall expand the parafermions \eqref{skdsjjds}, as well as the phase \eqref{skslshaoq1}, for $k\gg1$ and keep terms up to order $\nicefrac1k$.
This is a straightforward calculation which can be summarized in the following steps.
Firstly, we zoom $\th$ around zero as in \eqref{zooomm}. Then we express the variables $(\rho,\phi)$ in terms of $(y_1,y_2)$ by using \eqref{polar}.
Subsequently, we perform the field redefinition of \eqref{field.redefinition} keeping terms up to order $\nicefrac1k$.
Finally, we rewrite all expressions in terms of chiral coordinates, namely $x_\pm=x_1\pm ix_2$.
The end result is given by the following expressions \Big(we dismiss an overall factor of $\frac{1}{\sqrt{2k(1-\l^2)}}$\Big)
\be
\label{para.pert.re}
\begin{split}
&\Psi_+=\Big(\del x_-+\frac{F_+}{8k(1-\l^2)}\Big)\,\text{e}^{-i\phi_1}\,,\qquad
\Psi_-=\Big(\del x_+-\frac{F_-}{8k(1-\l^2)}\Big)\,\text{e}^{i\phi_1}\, ,
\\
&\bar\Psi_+=\Big(\bar\del x_+-\frac{\bar F_+}{8k(1-\l^2)}\Big)\text{e}^{-i\phi_1}\, ,\qquad
\bar\Psi_-=\Big(\bar\del x_-+\frac{\bar F_-}{8k(1-\l^2)}\Big)\,\text{e}^{i\phi_1}\,,
\end{split}
\ee
where
\begin{equation}
\label{Fscorrect}
\begin{split}
&F_+=(x_+^2-x_-^2)\left((1+\l^2)\del x_+-2\l\del x_-)\right)\, ,
\\
&
F_-=(x_+^2-x_-^2)\left((1+\l^2)\del x_--2\l\del x_+)\right)\, ,
\\
&\bar F_+=(x_+^2-x_-^2)\left((1+\l^2)\bar\del x_--2\l\bar\del x_+)\right)\, ,
\\
&
\bar F_-=(x_+^2-x_-^2)\left((1+\l^2)\bar\del x_+-2\l\bar\del x_-)\right)\  .
\end{split}
\end{equation}
Furthermore, the currents \eqref{JJc} have the following large $k$ expansion
\be
\label{currents.phase.pert}
\begin{split}
&J=\frac{i}{4k(1-\l^2)}\left(((1+\l^2)x_+-2\l x_-)\del x_--((1+\l^2)x_--2\l x_+)\del x_+\right)\,,
\\
&\bar J=\frac{i}{4k(1-\l^2)}\left(((1+\l^2)x_+-2\l x_-)\bar\del x_--((1+\l^2)x_--2\l x_+)\bar\del x_+\right)\,.
\end{split}
\ee
Note that the condition \eqref{skslshaoq1} determines $\phi_1$ provided that $F$ satisfies on-shell the condition
\begin{equation}\label{FbarF}
\del\bar F+\bar\del F
=-\frac{i\l}{k(1-\l^2)}(\del x_+\bar\del x_+-\del x_-\bar\del x_-)\,.
\end{equation}
Using the equation of motion \eqref{eqs2}, we find that to ${\cal O}(\nicefrac1k)$ \eqref{FbarF} is solved by\footnote{To  ${\cal O}(\nicefrac1k)$
the equations of motion \eqref{eqs2} read
$
\del\bar\del x_\pm=0\,.
$
}
\be
F=-\frac{i\l}{2k(1-\l^2)}(x_+\del x_+-x_-\del x_-)\, ,\quad
\ee
Employing the latter and \eqref{currents.phase.pert}, we find that
\be
\begin{split}
\label{currents.phase.pert.alt.final}
&J_F={i\ov 4 k}\frac{1+\l^2}{1-\l^2} \left(x_+ \del x_- - x_- \del x_+\right)\,,\\
&\bar J_F={i\ov 4 k}\frac{1+\l^2}{1-\l^2}  \left(x_+ \bar\del x_- - x_- \bar\del x_+\right)\,,
\end{split}
\ee
which generate $\phi_1$ through \eqref{phase1}. 
Let us note that the parafermions in \eqref{para.pert.re}, with 
the phase factor $\phi_1$ of \eqref{phase1} with \eqref{currents.phase.pert.alt.final}, are not on-shell chirally and anti-chirally conserved  for $\l\neq0$, since
\be
\bar\del\Psi_\pm=\mp i\,\del x_\mp\,\bar F\,\text{e}^{\mp i\phi_1}\,,\quad 
\del\bar\Psi_\pm=\pm i\,\bar\del x_\pm\, F\,\text{e}^{\mp i\phi_1}\,.
\ee
We close this subsection by noticing that in this set of variables the two symmetries of \eqref{sksldkdqq} are mapped respectively to
\be
\label{symmetry.pert}
\begin{split}
{\rm I}: \qq &\l\to\l^{-1}\,,\quad k\to-k\,,\\
{\rm II}:\qq & \l\to-\l\,,\quad x_\pm\to\pm ix_\pm\,,
\end{split}
\ee
under which the parafermions \eqref{para.pert.re} transform as
\be
\label{symme3}
\begin{split}
{\rm I}:\qq & \Psi_\pm\to\Psi_\pm\,,\quad \bar\Psi_\pm\to\,\bar\Psi_\pm\,,\\
{\rm II}:\qq  &\Psi_\pm\to\mp i\,\Psi_\pm\,,\quad \bar\Psi_\pm\to\pm i\,\bar\Psi_\pm\,,
\end{split}
\ee
while \eqref{currents.phase.pert} and \eqref{currents.phase.pert.alt.final} remain intact. 
In the computations which follow we shall make use of \eqref{para.pert.re}, with 
the phase factor $\phi_1$ of \eqref{phase1}, \eqref{currents.phase.pert.alt.final}, keeping terms up to order ${\cal O}(\nicefrac1k)$.

\subsection{The anomalous dimension of parafermion bilinear}
\label{bilin}

The aim of this subsection is to calculate the anomalous dimension of the operator that perturbs the $SU(2)/U(1)$ CFT.
This operator is bilinear in the parafermions and its form is given by the classical parafermions bilinear \cite{Sfetsos:2013wia}
\be
\label{pertparaf}
{\cal O} =\Psi_+ \bar \Psi_- + \Psi_-\bar \Psi_+ \, ,
\ee
where we note that  the phase factors cancel, as one may readily verify using \eqref{claparaf}.

\no
Taking the large $k$-limit we obtain that the leading term of $\cal O$
is given by $2(\del x_1 \bar \del x_1 - \del x_2 \bar \del x_2)$.
Had we kept the leading correction, a  bilinear in the parafermions operator
would have been of the generic form (we ignore the irrelevant factor
of $2$)
\be
\label{bibi}
{\cal O} =\del x_1 \bar \del x_1 - \del x_2 \bar \del x_2
+ {1\ov k} \sum_{i=1}^4 c_i {\cal O}_i + \dots \  ,
\ee
for some coefficients $c_i$ and a basis of operators with engineering dimension $(1,1)$  ${\cal O}_i$.\footnote{Such a basis
is
\begin{equation*}
\begin{split}
{\cal O}_1 = x_1 x_2 \del x_1\bar \del x_2\,,\quad {\cal O}_2 =x_1 x_2 \del x_2 \bar \del x_1\,,\quad
{\cal O}_3 =\ha x_1^2 \del x_2 \bar \del x_2\,,\quad  {\cal O}_4 = \ha x_2^2 \del x_1 \bar \del x_1\,.
\end{split}
\end{equation*}
}
However, notice that the overlap of the leading and subleading term in \eqref{bibi} is zero. This means that one can safely ignore the subleading term in \eqref{bibi}, at least up to ${\cal O}(\nicefrac1k)$ which is the order we are working in the present paper.

\no
In order to proceed, we will also need the interaction terms of the action \eqref{acctt10} up to order ${\cal O}(\nicefrac1k)$.
There is a single such term which in the Euclidean regime reads
\be
\begin{split}
\label{sintt}
S_{\rm int} & = {\hat g_{12}\ov 2\pi k} \int \text{d}^2z\, x_1 x_2 (\del x_1 \bar \del x_2 + \del x_2\bar \del x_1)
\\
& = -{\hat g_{12}\ov 16 \pi k} \int \text{d}^2z\,  (x_+^2 - x_-^2)(\del x_+ \bar \del x_+ - \del x_- \bar \del x_-)  \ .
\end{split}
\ee

\no
The two-point function then reads
\be\label{OO}
\begin{split}
&
\langle {\cal O}(z_1,\z_1) {\cal O}(z_2,\z_2)  \rangle = {2\ov |z_{12}|^4}
-{\hat g_{12}\ov 2\pi k} \int \text{d}^2z\, \langle \big(\del x_1(z_1) \bar \del x_1(\z_1) - \del x_2(z_1) \bar \del x_2(\z_1)\big)
\\
&\qq\quad \times \big(\del x_1(z_2) \bar \del x_1(z_2) - \del x_2(z_2) \bar \del x_2(\z_2)\big)
\big(x_1 x_2 (\del x_1 \bar \del x_2 + \del x_2 \bar \del x_1)\big)(z,\z)\rangle
\\
& \qq\qq\quad\
=
{2\ov |z_{12}|^4}  +2\, {\hat g_{12}\ov 2\pi k} (I_1 + I_2) \ ,
\end{split}
\ee
where the minus sign in the second term above, corresponding to a single insertion of the interaction term, is due to
the fact that in the Euclidean regime we have the term $e^{-S_{\rm int}}$ in the correlators.
By inspecting \eqref{OO} it is straightforward to see   that we need two kinds of contractions, namely
\be
\begin{split}
 I_1 = &\int \text{d}^2z\, \langle \del x_1(z_1) \bar \del x_1(\z_1) \del x_1(z) x_1(z,\z)\rangle
\langle \del x_2(z_2) \bar \del x_2(\z_2) \bar \del x_2(\z) x_2(z,\z)\rangle
\\
= & \int {\text{d}^2z\ov (z-z_1)^2 (z-z_2) (\z-\z_1) (\z-\z_2)^2}
\end{split}
\ee
and
\be
\begin{split}
I_2 = &  \int \text{d}^2z\, \langle \del x_1(z_1) \bar \del x_1(\z_1) \bar \del x_1(\z) x_1(z,\z)\rangle
\langle \del x_2(z_2) \bar \del x_2(\z_2) \del x_2(z) x_2(z,\z)\rangle
\\
= & \int {\text{d}^2z\ov (z-z_1) (z-z_2)^2 (\z-\z_1)^2 (\z-\z_2)}\, .
\end{split}
\ee
Using \eqref{first} twice we find that
\be
\langle {\cal O}(z_1,\z_1) {\cal O}(z_2,\z_2) \rangle = {2\ov |z_{12}|^4}\left(1+ {2\hat g_{12}\ov k}\left(1+\!\ln\frac{\varepsilon^2}{|z_{12}|^2}\right)\right)\,,
\ee
from which we read the anomalous dimension of the parafermion bilinear to be
\be
\label{skfqoqwlql}
\g^{{\cal O}} ={2\hat g_{12}\ov k}  =  -{2 \ov k}{1+\l^2\ov 1-\l^2}\ .
\ee
 This expression is in perfect agreement with equation (4.16) of \cite{Georgiou:2019nbz},
 where the anomalous dimension of the
 perturbing operator was found by using the geometry in the space of couplings. Notice that in the conformal point the anomalous
 dimension of the composite operator is twice the
anomalous dimension of the holomorphic (anti-holomorphic ) parafermion 
which equals $-\nicefrac1k$ \cite{Fateev:1985mm}.

The reader may wonder if there are other operators with the same engineering dimension as ${\cal O}$ which may mix with it.
The  operators of such equal engineering dimension are
\be
\begin{split}
\label{bibi1}
\tilde {\cal  O} =\del x_1 \bar \del x_1 + \del x_2 \bar \del x_2 + {\cal O}(\nicefrac1k)\,,\quad 
\tilde {\cal O}_\pm =\del x_1 \bar \del x_2 \pm\, \del x_2 \bar \del x_1 + {\cal O}(\nicefrac1k)\ ,
\end{split}
\ee
where the corrections are of same form with those for ${\cal O}$ in \eqref{bibi}.
It turns out that these operator does not mix with ${\cal O}$ and among themselves at the free field point and also at ${\cal O}(\nicefrac1k)$.
Moreover, one may easily show that $\tilde {\cal O}$ has the opposite anomalous dimension as that in \eqref{skfqoqwlql}.
The anomalous dimension of $\tilde {\cal O}_\pm$ is computed for completeness in appendix \ref{appA}.

\subsection{Anomalous dimension of the single parafermion}\label{singlepar}

The goal of this subsection is to compute, using the free field expansion, the two-point functions of the parafermions to order  $\nicefrac1k$
 from which one can read the anomalous dimension of the deformed parafermion.

\no
We first consider the correlator
$
\langle \Psi_+\Psi_+\rangle\, .
$
This vanishes at the conformal point since it is not neutral. Employing the expressions above we find that
\be\label{psi +psi+}
\begin{split}
&\langle \Psi_+(z_1,\z_1)\Psi_+(z_2,\z_2)\rangle=\\
&\quad +{\hat g_{12}\ov 16\pi k}\int\text{d}^2z\langle \del x_-(z_1) \del x_-(z_2)
 \big((x_+^2-x_-^2) (\del x_+ \bar \del x_+ - \del x_- \bar \del x_-)\big)(z,\z)\rangle\\
&\quad +\frac{1}{8k(1-\l^2)}\left(\langle F_+(z_1,\bar z_1)\del x_-(z_2) \rangle+\langle\del x_-(z_1)F_+(z_2,\bar z_2) \rangle\right)\\
&\quad -i\langle \del x_-(z_1)\del x_-(z_2)
\phi_1(z_1,z_0)\rangle
-i\langle \del x_-(z_1)\del x_-(z_2)
\phi_1(z_2,z_0)\rangle=0\,,
\end{split}
\ee
where for the phase $\phi_1$ we will use the leading ${\cal O}(\nicefrac1k)$ of  \eqref{phase1} with  \eqref{currents.phase.pert.alt.final} so that
the corresponding terms in \eqref{psi +psi+} are indeed of ${\cal O}(\nicefrac1k)$. The second of \eqref{psi +psi+} involves the interaction term of the action and vanishes
due to the fact that the interaction term is normal ordered. The third and fourth line of \eqref{psi +psi+} originate from the ${\cal O}(\nicefrac1k)$ corrections to the parafermion operators and from the phase $\phi_1$, respectively. They both vanish either because in the process one necessarily encounters propagators of the form $\langle x_+ \,x_+ \rangle$ or $\langle x_- \,x_- \rangle$ which are identically zero (terms in the fourth line) or because of normal ordering (terms in the third line).

\no
We now turn to the neutral two point function
$
\langle \Psi_+ \Psi_- \rangle\,.
$
To ${\cal O}(\nicefrac1k)$ this correlator equals to
\be\label{+-}
\begin{split}
&\langle \Psi_+(z_1,\z_1)\Psi_-(z_2,\z_2)\rangle=\\
&\quad -\frac{2}{z_{12}^2}+\frac{1}{8k(1-\l^2)}\left(\langle F_+(z_1,\z_1) \del x_+(z_2)\rangle-\langle \del x_-(z_1) F_-(z_2,\z_2)\rangle\right)
\\
&\quad + {\hat g_{12}\ov 16\pi k}\int\text{d}^2z\langle \del x_-(z_1) \del x_+(z_2)
\big((x_+^2-x_-^2) (\del x_+ \bar \del x_+ - \del x_- \bar \del x_-)\big)(z,\z)\rangle \\
&\quad -i\langle \del x_-(z_1)\del x_+(z_2)\phi_1(z_1,z_0)\rangle
+i\langle \del x_-(z_1)\del x_+(z_2)\phi_1(z_2,z_0)\rangle\,.
\end{split}
\ee
In the above expression, apart from the free part, only the last two terms related to the phase factor in the parafermions are
non-vanishing and equal to
\be\label{psi+psi-}
\begin{split}
&\frac{z_{12}}{k}\,\frac{1+\l^2}{1-\l^2} \bigg(-\int_{z_0}^{z_1}\text{d}z+\int_{z_0}^{z_2}\text{d}z \bigg)\frac{1}{(z-z_1)^2(z-z_2)^2}
\\
&=\frac{1}{z_{12}^2}\frac{2}{k}\frac{1+\l^2}{1-\l^2} \bigg(1+\ln\frac{\varepsilon^2}{z_{12}^2}\bigg)
+ \frac{2}{k}\,\frac{1+\l^2}{1-\l^2}\frac{1}{\varepsilon\, z_{12}}\ .
\end{split}
\ee
All other terms in \eqref{psi+psi-} vanish due to the fact that both the interaction term and $F_\pm$ are normal ordered.
In evaluating the line integrals of \eqref{psi+psi-} we have introduced a small distance cut-off $\varepsilon$ so that the integration point never coincides with the end points of the integration. Therefore, we find that
\be
\label{anom}
\langle \Psi_+(z_1,\z_1)\Psi_-(z_2,\z_2)\rangle=-\frac{2}{z_{12}^2}\Bigg(1-\frac{1}{k}\frac{1+\l^2}{1-\l^2}
\Big(1+\ln\frac{\varepsilon^2}{z_{12}^2}\Big)\Bigg) + \frac{2}{\varepsilon\,k}\,\frac{1+\l^2}{1-\l^2}\frac{1}{z_{12}}\, .
\ee
The $\nicefrac{1}{\varepsilon}$ pole can be absorbed by a field redefinition of the $\Psi_\pm$'s, namely $\Psi_\pm \rightarrow \Psi_\pm+\d\Psi_\pm$, where
\be\label{redef}
\d\Psi_\pm=-\frac{f_\pm}{\varepsilon \,k}x_\mp\,,\quad f_+-f_-=\frac{1+\l^2}{1-\l^2}\,.
\ee
Note that we do not need to specify both parameters $f_\pm$, just their difference as above and in addition this relation is
invariant under the symmetries \eqref{symme3}. At $\l=0$, the correlator \eqref{psi+psi-} is in agreement with 
the CFT result to leading order in $\nicefrac1k$ after one makes the  the following rescaling
\be
\label{matchCFT}
\Psi=i\sqrt{2}\left(1-\frac{1}{2k}\right)\Psi_\text{CFT}\, .
\ee
Finally, from \eqref{anom} one can read the anomalous dimension of $\Psi$ which is given by
\be
\gamma_\Psi=-\frac{1}{k}\frac{1+\l^2}{1-\l^2}\,.
\ee
Notice that $\gamma_\Psi$ is half the anomalous dimension of the composite operator \eqref{skfqoqwlql}.
At the conformal point it is indeed,
the deviation from unity of the holomorphic dimension (equal to the anti-holomorphic one) of the parafermion which,
as already noted, equals $1-\nicefrac1k$.
For completeness, we note that the conjugate correlator $\langle \bar\Psi_+\bar\Psi_-\rangle$ takes the
form \eqref{anom} as well.

\subsection{Four-point correlation functions}
\label{corfun}

In this section, we calculate four-point correlators of parafermions.
To begin with, note that at the CFT point correlation functions with an odd number of parafermions vanish.
In the following we will argue that the same is true for the deformed model as well. Indeed, to ${\cal O}(\nicefrac1k)$ the number of fields involved in the
correlator is odd since the phase may contribute two and the insertion from the interacting
part of the action four fields. Any correlator with an odd number of free fields vanishes.
This argument should hold to all orders in the large $k$ expansion since every extra power of $\nicefrac1k$
contributes two more fields in the correlators.

\no
At the CFT point correlators with odd number of holomorphic (or antiholomorphic) parafermions also  vanish identically
due to charge conservation. However, CFT point correlators with an even number of holomorphic (or antiholomorphic)
parafermions and neutral do not vanish. Away from the CFT point  correlators with an even number of fields receive corrections.
Using conformal perturbation one can easily convince oneself that the corrections come to even order in the $\lambda$ expansion.

\subsubsection{Correlation function $\langle \Psi_+  \Psi_- \Psi_+ \Psi_- \rangle $}
\label{4pt1}

Consider, as our first example, the four-point function
\ba
\label{+_+_}
&&\langle \Psi_+(z_1,\z_1)  \Psi_-(z_2,\z_2) \Psi_+(z_3,\z_3) \Psi_-(z_4,\z_4)\rangle
= {4\ov z_{12}^2 z_{34}^2} +  {4\ov z_{14}^2 z_{23}^2}
\nonumber
\\
&& +\frac{1}{8k(1-\l^2)}\left(\langle F_+(z_1,\z_1)  \del x_+(z_2) \del x_-(z_3) \del x_+(z_4)
\rangle +\langle F_+(z_3,\z_3)  \del x_+(z_2) \del x_-(z_1) \del x_+(z_4)
\rangle \right.
\nonumber
\\
&& \left.-\langle F_-(z_2,\z_2)\del x_-(z_1)  \del x_-(z_3)
\del x_+(z_4)  \rangle -\langle F_-(z_4,\z_4)\del x_-(z_1)\del x_+(z_2)   \del x_-(z_3)
 \rangle\right)
\\
&& -i\langle\phi_1(z_1,z_0) \del x_-(z_1) \del x_+(z_2)   \del x_-(z_3)
\del x_+(z_4)\rangle
-i\langle\phi_1(z_3,z_0) \del x_-(z_1) \del x_+(z_2)   \del x_-(z_3)
\del x_+(z_4)\rangle
\nonumber
\\
&&+i\langle\phi_1(z_2,z_0)\del x_-(z_1) \del x_+(z_2)   \del x_-(z_3)
\del x_+(z_4) \rangle +i\langle\phi_1(z_4,z_0)\del x_-(z_1) \del x_+(z_2)   \del x_-(z_3)
\del x_+(z_4) \rangle
\nonumber
\\
 &&  +{\hat g_{12}\ov 16\pi k}
 \int\text{d}^2z\langle\del x_-(z_1) \del x_+(z_2)   \del x_-(z_3)  \del x_+(z_4)
 \big((x_+^2-x_-^2) (\del x_+ \bar \del x_+ - \del x_- \bar \del x_-)\big)(z,\z)\rangle\ .
\nonumber
\ea
To calculate this is a rather tedious but straightforward task. To start, let us briefly describe the various contributions.
The second and third line of \eqref{+_+_}, originate from the $\nicefrac1k$ corrections of $\Psi$'s \eqref{Fscorrect} and involve only free contractions, straightforwardly yielding
\be
\frac{2}{k}\frac{1+\l^2}{1-\l^2}\left(\frac{1}{z_{14}z_{34}z_{24}^2}+
\frac{1}{z_{12}z_{14}z_{13}^2}-\frac{1}{z_{12}z_{23}z_{24}^2}-\frac{1}{z_{23}z_{24}z_{12}^2}\right)\,.
\ee
The fourth and fifth line, which originate from the phase $\phi_1$, give after performing ordinary integrations over $z$ the following contribution
\be
\begin{split}
&-\frac{8}{k}\frac{1+\l^2}{1-\l^2}\left(\frac{1}{z_{12}^2z_{34}^2}+\frac{1}{z_{14}^2z_{23}^2}\right)
+\frac{8}{k}\frac{1+\l^2}{1-\l^2}\frac{1}{z_{12}z_{23}z_{14}z_{34}}\\
&+\frac{8}{k}\frac{1+\l^2}{1-\l^2}\left(\frac{1}{z_{12}^2z_{34}^2}+\frac{1}{z_{14}^2z_{23}^2}\right)\ln\frac{z_{12}z_{23}z_{14}z_{34}}{z_{13}z_{24}}\,.
\end{split}
\ee
Finally, from the last line we find
\be
\frac{2\hat g_{12}}{k}\left(\frac{1}{z_{14}z_{34}z_{24}^2}+\frac{1}{z_{12}z_{14}z_{13}^2}-\frac{1}{z_{12}z_{23}z_{24}^2}-\frac{1}{z_{23}z_{24}z_{12}^2}\right)\,.
\ee
This result is easily obtained since the integration over $z$ is immediately performed since the free contractions necessarily give
contact terms of the form $\d^{(2)}(z-z_i)$.

Adding the various contributions we finally find
\be
\begin{split}
&\langle\Psi_+(z_1,\z_1)\Psi_-(z_2,\z_2)\Psi_+(z_3,\z_3)\Psi_-(z_4,\z_4)\rangle=4\frac{1-\frac{2}{k}\frac{1+\l^2}{1-\l^2}}{z_{12}^2z_{34}^2}
+4\frac{1-\frac2k\frac{1+\l^2}{1-\l^2}}{z_{14}^2z_{23}^2}\\
&\quad +\frac{8}{k}\frac{1+\l^2}{1-\l^2}\frac{1}{z_{12}z_{23}z_{14}z_{34}}
+\frac{8}{k}\frac{1+\l^2}{1-\l^2}\left(\frac{1}{z_{12}^2z_{34}^2}+\frac{1}{z_{14}^2z_{23}^2}\right)\ln\frac{z_{12}z_{23}z_{14}z_{34}}{z_{13}z_{24}}\,.
\end{split}
\ee
At $\l=0$, this is in agreement with the CFT result \cite{Fateev:1985mm} after the rescaling \eqref{matchCFT} and of course 
in the large $k$ expansion up to ${\cal O}(\nicefrac1k)$.
Furthermore, this result is in agreement with perturbation which predicts that corrections to the correlator under consideration
should start at order $\l^2$. This is so because the contribution linear in $\lambda$ will involve a single anti-holomorphic parafermion and will, thus, be vanishing.
Lastly, we would like to comment on the efficiency of the method initiated in \cite{Georgiou:2019aon} and further developed here. Our method provides the
exact in the deformation parameter $\lambda$ correlators in contradistinction to the usual conformal perturbation theory which only gives results order by 
order in the $\lambda$ expansion.
The power of our approach resides on the fact that the exact in $\lambda$ effective action (see \eqref{actioncoset}) is at hand.

\subsubsection{Correlation function  $\langle \Psi_+  \Psi_+\bar \Psi_-\bar  \Psi_- \rangle$}
\label{4pt2}

Our second four-point correlation function has the following large $k$ expansion
\ba
&&\langle \Psi_+(z_1,\z_1)  \Psi_+(z_2,\z_2) \bar \Psi_-(z_3,\z_3) \bar\Psi_-(z_4,\z_4)\rangle
  \nonumber
 \\
&&=
\frac{1}{8k(1-\l^2)}\left(\langle F_+(z_1,\z_1) \del x_-(z_2)   \bar\del x_-(\z_3) \bar\del x_-(\z_4) \rangle +
\langle F_+(z_2,\z_2) \del x_-(z_1)   \bar\del x_-(\z_3) \bar\del x_-(\z_4) \rangle\right.
 \nonumber
 \\
&&
\left.+\langle \bar F_-(z_3,\z_3) \del x_-(z_1)
\del x_-(z_2)   \bar\del x_-(\z_4)  \rangle +\langle \bar F_-(z_4,\z_4) \del x_-(z_1)
\del x_-(z_2)   \bar\del x_-(\z_3)  \rangle\right)
 \\
&&
-i\langle\phi_1(z_1,z_0)\del x_-(z_1)  \del x_-(z_2)  \bar\del x_-(\z_3)   \bar\del x_-(\z_4)\rangle
-i\langle\phi_1(z_2,z_0)\del x_-(z_1)  \del x_-(z_2)  \bar\del x_-(\z_3)   \bar\del x_-(\z_4)\rangle
 \nonumber
 \\
&&
+i\langle\phi_1(z_3,z_0)
\del x_-(z_1)  \del x_-(z_2)  \bar\del x_-(\z_3)   \bar\del x_-(\z_4)\rangle
+i\langle\phi_1(z_4,z_0)
\del x_-(z_1)  \del x_-(z_2)  \bar\del x_-(\z_3)   \bar\del x_-(\z_4)\rangle
  \nonumber \\
 &&+ {\hat g_{12}\ov 16\pi k} \int \text{d}^2z\,
 \langle \del x_-(z_1) \del x_-(z_2) \bar \del x_-(z_3)  \bar \del x_-(\z_4)
 \big((x_+^2 - x_-^2)(\del x_+ \bar \del x_+ - \del x_- \bar \del x_-)\big)(z,\z)\rangle \ .
  \nonumber
\ea
A non-vanishing correlator requires equal number of $x_+$ and $x_-$'s. Clearly, only the last line may contribute, giving
\ba
\label{dfhu1}
&& \langle \Psi_+(z_1,\z_1)  \Psi_+(z_2,\z_2) \bar \Psi_-(z_3,\z_3) \bar\Psi_-(z_4,\z_4)\rangle
\nonumber
\\
&& =  {\hat g_{12}\ov 16\pi k} \int \text{d}^2z
 \langle \del x_-(z_1) \del x_-(z_2) \bar \del x_-(\z_3)  \bar \del x_-(\z_4)  (x_+^2 \del x_+ \bar \del x_+ )(z,\z)\rangle
 \nonumber \\
 && = {\hat g_{12}\ov \pi k}  \int \text{d}^2z\, \bigg( {1\ov (z-z_1)(z-z_2)^2 (\z-\z_3)(\z-\z_4)^2} + {1\ov (z-z_1)^2(z-z_2) (\z-\z_3)(\z-\z_4)^2}
 \nonumber  \\
& & + {1\ov (z-z_1)(z-z_2)^2 (\z-\z_3)^2(\z-\z_4)}  + {1\ov (z-z_1)^2(z-z_2) (\z-\z_3)^2(\z-\z_4)}\bigg )\,.
\ea
To evaluate the above integrals, we use \eqref{smslsdldsm} four times with appropriate renaming of
the $z_i$'s. Adding them up we find a vanishing result
\be
\langle \Psi_+(z_1,\z_1)\Psi_+(z_2,\z_2)\bar \Psi_-(z_3,\z_3)\bar \Psi_-(z_4,\z_4)\rangle=0\,.
\ee

\subsubsection{Correlation function  $\langle \Psi_+  \Psi_- \bar\Psi_+\bar \Psi_- \rangle$ }
\label{4pt3}

The third four-point correlator that we will consider is the following
\ba
\label{+-b+b-}
&&\langle \Psi_+(z_1,\z_1)  \Psi_-(z_2,\z_2) \bar \Psi_+(z_3,\z_3) \bar\Psi_-(z_4,\z_4)\rangle
 = {4\ov z_{12}^2 \z_{34}^2}
\nonumber\\
&&+  \frac{1}{8k(1-\l^2)}\left(\langle F_+(z_1,\z_1)  \del x_+(z_2)  \bar \del x_+(\z_3)
\bar \del x_-(\z_4) \rangle -\langle F_-(z_2,\z_2) \del x_-(z_1) \bar\del x_+(\z_3)
\bar\del x_-(\z_4) \rangle \right.
\nonumber\\
&&\left.-\langle \bar F_+(z_3,\z_3) \del x_-(z_1)  \del x_+(z_2)
 \bar \del x_-(\z_4) \rangle
 + \langle \bar F_-(z_4,\z_4) \del x_-(z_1)  \del x_+(z_2)
 \bar\del x_+(\z_3)   \rangle \right)
 \\
&&\quad-i\langle\phi_1(z_1,z_0)\del x_-(z_1) \del x_+(z_2)
\bar \del x_+(\z_3)  \bar \del x_-(\z_4)  \rangle
+i\langle\phi_1(z_2,z_0)\del x_-(z_1) \del x_+(z_2)
\bar \del x_+(\z_3)  \bar \del x_-(\z_4)  \rangle
\nonumber\\
&&
\quad
-i\langle\phi_1(z_3,z_0) \del x_-(z_1) \del x_+(z_2)
\bar \del x_+(\z_3)  \bar \del x_-(\z_4)  \rangle
+i\langle\phi_1(z_4,z_0) \del x_-(z_1) \del x_+(z_2)
\bar \del x_+(\z_3)  \bar \del x_-(\z_4)  \rangle
\nonumber\\
 && \quad
 + {\hat g_{12}\ov 16\pi k} \int \text{d}^2 z
 \langle  \del x_-(z_1) \del x_+(z_2) \bar \del x_+(\z_3)  \bar \del x_-(\z_4)
 \big((x_+^2 - x_-^2)(\del x_+ \bar \del x_+ - \del x_- \bar \del x_-)\big)(z,\z)\rangle \ .
\nonumber
\ea
To evaluate \eqref{+-b+b-} is a rather cumbersome  but straightforward computation. Let us briefly describe the various contributions.
The second and third line, related to the $\nicefrac1k$ corrections of the parafermions $\Psi$'s \eqref{Fscorrect}, trivially vanish once we ignore contact terms involving only external points.
The fourth and fifth lines, that are related to the phase
$\phi_1$ of $\Psi$'s, yield  after performing ordinary integrations the following result
\be
\begin{split}
&\frac{1}{k}\frac{1+\l^2}{1-\l^2}\frac{4}{z_{12}^2\z_{34}^2}\left(-2-\ln\frac{\varepsilon^2}{z_{12}^2}-\ln\frac{\varepsilon^2}{\bar z_{34}^2}+
\ln\frac{|z_{23}|^2|z_{14}|^2}{|z_{13}|^2|z_{24}|^2}\right.\\
&\left.+\frac12\left(\frac{z_{12}z_{34}}{z_{24}z_{23}}+
\frac{z_{12}z_{34}}{z_{13}z_{14}}+\frac{\z_{12}\z_{34}}{\z_{23}\z_{13}}+\frac{\z_{12}\z_{34}}{\z_{14}\z_{24}}\right)\right)\,.
\end{split}
\ee
Finally, the last line of \eqref{+-b+b-} originating from the interaction insertion equals to
\be
{4\ov k}\frac{\hat g_{12}}{z_{12}^2\z_{34}^2}\left(\ln\frac{|z_{23}|^2|z_{14}|^2}{|z_{13}|^2|z_{24}|^2}+\frac12\left(\frac{z_{12}z_{34}}{z_{24}z_{23}}+
\frac{z_{12}z_{34}}{z_{13}z_{14}}+\frac{\z_{12}\z_{34}}{\z_{23}\z_{13}}+\frac{\z_{12}\z_{34}}{\z_{14}\z_{24}}\right)\right)\,.
\ee
In order to obtain the last result, we have used twice \eqref{smslsdldsm} with an appropriate renaming of
the $z_i$'s.

\no
Adding all the  above contributions we find that
\begin{equation}
\begin{split}
\langle \Psi_+(z_1,\z_1)\Psi_-(z_2,\z_2)\bar \Psi_+(z_3,\z_3) &\bar \Psi_-(z_4,\z_4)\rangle=
\frac{4}{z_{12}^2\bar z_{34}^2}
\Bigg(1-\frac{2}{k}\frac{1+\l^2}{1-\l^2}
\\
&
-\frac{1}{k}\frac{1+\l^2}{1-\l^2}\bigg(\ln\frac{\varepsilon^2}{z_{12}^2}+\ln\frac{\varepsilon^2}{\bar z_{34}^2}\bigg)\Bigg)\,.
\end{split}
\end{equation}
At $\l=0$, this result is under the rescaling \eqref{matchCFT} in agreement with the CFT result which enforces factorization of the four-point function into two-point functions holomorphic and anti-holomorphic, respectively.

\no
\subsubsection{Correlation function  $ \langle \Psi_+  \Psi_+ \Psi_+ \Psi_- \rangle$}
\label{4pt4}

The only contribution for this correlator comes from the free contractions involving the $\nicefrac1k$ terms in the expression for the parafermions $\Psi$'s \eqref{para.pert.re} and \eqref{Fscorrect}, yielding finally
\ba
&& \langle\Psi_+(z_1,\z_1)\Psi_+(z_2,\z_2)\Psi_+(z_3,\z_3)\Psi_-(z_4,\z_4)\rangle=  \frac{4\l}{k(1-\l^2)}\Bigg(\frac{1}{z_{14}^2z_{12}z_{13}}-\frac{1}{z_{24}^2z_{12}z_{23}}
\nonumber
\\
&&\qq\qq
+ \frac{1}{z_{34}^2z_{23}z_{13}}
 -\frac{1}{z_{14}^2z_{24}z_{34}}-\frac{1}{z_{24}^2z_{14}z_{34}}-\frac{1}{z_{34}^2z_{14}z_{24}}\Bigg)=0\,.
\ea
We should mention that the result above is consistent with perturbation theory around the conformal point which predicts that it is exactly zero to all orders in the $\nicefrac1k$ expansion since the correlator at hand is not a neutral one.

\subsubsection{ Correlation function $\langle \Psi_+  \Psi_+ \Psi_+ \Psi_+ \rangle$}
\label{4pt5}

There are two contributions for this correlator.
The first one comes from the $\nicefrac1k$ terms of the  $\Psi$'s \eqref{para.pert.re} and  \eqref{Fscorrect}, it involves only free contractions and reads
\begin{equation}
\label{jjowo1}
\begin{split}
\frac{1+\l^2}{8k(1-\l^2)}  \Big(&\langle(x_+^2\del x_+)(z_1,\z_1)\del x_-(z_2)\del x_-(z_3)\del x_-(z_4)\rangle
\\
&
 +\langle(x_+^2\del x_+)(z_2,\z_2)\del x_-(z_1)\del x_-(z_3)\del x_-(z_4)\rangle
\\
&
+\langle(x_+^2\del x_+)(z_3,\z_3)\del x_-(z_1)\del x_-(z_2)\del x_-(z_4)\rangle
\\
&+\langle(x_+^2\del x_+)(z_4,\z_4)\del x_-(z_1)\del x_-(z_2)\del x_-(z_3)\rangle\Big)\,.
\end{split}
\ee
The second contribution comes from the interaction insertion and is given by
\be
\frac{\hat g_{12}}{16\pi k}\int\text{d}^2z\,\langle\del x_-(z_1)\del x_-(z_2)\del x_-(z_3)\del x_-(z_4)(x_+^2\del x_+\bar\del x_+)(z,\z)\rangle\,.
\ee
To evaluate this integral we contract $\bar\del x_+$ with the $(\del x_-)_i$'s yielding contact terms proportional to $\d^{(2)}(z-z_i)$. Evaluating then the integral we obtain
an expression with free contractions among external point only,
which is, after substituting $\hat g_{12}$ from \eqref{coopl2}, is precisely the opposite of that in \eqref{jjowo1}.
Combining these two contributions we find the vanishing result
\be
\langle\Psi_+(z_1,\z_1)\Psi_+(z_2,\z_2)\Psi_+(z_3,\z_3)\Psi_+(z_4,\z_4)\rangle=0\,.
\ee
As in the previous correlation function this result is in full agreement with perturbation theory around the conformal point and vanishes to all orders in the 
$\nicefrac1k$ expansion.

A couple of important comments are in order. The first one concerns the $\nicefrac1\varepsilon$ poles in the calculation of the four-point functions. These infinite contributions have been suppressed in the presentation
because on can absorb them by employing  precisely the same redefinition \eqref{redef} used in the two-point function \eqref{+-}.
The second comment concerns the form of the two and four-point correlators. By just inspecting the results one can see that they behave as if the theory was conformal albeit with a redefined level $\tilde k=k{1-\l^2\ov 1+\l^2}$.
In particular, this implies that the $\lambda$-dressed parafermions of the deformed theory will satisfy the same Poisson bracket algebra as the one satisfied by the parafermions at the conformal point but with $\tilde k$ in the place of $k$.

\section{Discussion and future directions}
\label{sec5}

We studied the $\lambda$-deformed $SU(2)/U(1)$ $\sigma$-model by using a free field expansion instead of conformal perturbation and the underlying parafermionic algebras.
Expanding along the lines of \cite{Georgiou:2019aon} the perturbation is organized as a series expansion for large values of $k$.
This approach has the advantage that all deformation effects are fully encoded in the coupling constant coefficients and in the form of the operators.  
This prescription allowed for the computation the RG flow of the deformation parameter $\lambda$ and the anomalous dimension of the parafermion bilinear perturbation, driving the theory away from the conformal point, as  exact functions of $\lambda$ and up to ${\cal O}(\nicefrac1k)$. Then, we introduced the corresponding $\lambda$-dressed parafermions containing a non-local phase in their expressions -- which ensures on-shell chirality at the conformal point.  Subsequently, we evaluate their anomalous dimensions and their four-point functions, odd-point correlation functions identically vanish, again to all-orders in $\lambda$ and up to ${\cal O}(\nicefrac1k)$. 
The derived results are in agreement with CFT expectations at $\l=0$ and are invariant under the duality-type symmetries \eqref{sksldkdqq}.

\no
The $SU(2)/U(1)$ is the simplest possible symmetric coset space. It should be possible to study other $\lambda$-deformed coset 
CFT based on symmetric spaces using free fields as a basis. 
Furthermore, it was argued in \cite{Georgiou:2019nbz} that 
the $\lambda$-deformed $SU(2)\times SU(2)/U(1)$ and $SU(2)/U(1)$ CFTs are closely 
related since they share the same symmetries and $\beta$-function to all-orders in $\lambda$ and $k$. It would be interesting to check and understand this relation further using the techniques of the present work. 

\no
Another interesting direction is to extend the current set for deformations of non-symmetric coset CFTs. 
A class of such integrable models is the $\lambda$-deformed $G_{k_1}\times G_{k_2}/G_{k_1+k_2}$ coset CFTs, 
which, assuming that $k_1>k_2$, flows in the infrared  to the  $G_{k_2}\times G_{k_1-k_2}/G_{k_1}$ coset CFT
\cite{Sfetsos:2017sep}. These models have a richer parameter space and as a result a very interesting
generalization of the duality-type symmetries \eqref{sksldkdqq} (see eq. (2.17)  in that work) which dictates their behavior and which reduces to 
the first one in \eqref{sksldkdqq} for equal levels. In addition, in these models the perturbation is driven by non-Abelian parafermion bilinears \cite{Bardakci:1990ad} which are much less understood than their Abelian counterparts \cite{Bardacki:1990wj} we utilized in the present work.
We believe that our free field techniques will be instrumental in understanding these type of non-symmetric coset CFTs.

The present setup may be applied in the study of the two-parameter integrable deformations of symmetric coset spaces constructed in \cite{Sfetsos:2015nya}. 
Such type of deformations are applicable to a restricted class of symmetric spaces provided that they satisfy a certain gauge invariance condition involving a non-trivial solution of the Yang--Baxter equation (see Eq.(5.19) in that work). Examples include the $SU(2)/U(1)$ \cite{Sfetsos:2015nya}, the $SO(N+1)/SO(N)$ \cite{Lunin:2017bha} and the recently studied $\mathbb{C}P^n$ models \cite{Demulder:2020dlo}.
In the aforementioned examples, this second parameter can be set to zero via parameter redefinitions and/or diffeomorphisms. Such redefinitions come with caveats due to the topological nature of $k$ and the non-trivial monodromies that the associated parafermions will presumably have. Therefore, a proper study of this class of models,  in the presence context in terms of free fields, should include both deformation parameters.

\subsection*{Acknowledgments}

The work of G. Georgiou and K. Siampos on this project has received funding from the Hellenic Foundation for Research and Innovation (H.F.R.I.) 
and the General Secretariat for Research and Technology (G.S.R.T.), under grant
agreement No 15425. \\
The research work of K. Sfetsos was supported by the Hellenic Foundation for
Research and Innovation (H.F.R.I.) under the ``First Call for H.F.R.I.
Research Projects to support Faculty members and Researchers and
the procurement of high-cost research equipment grant'' (MIS 1857, Project Number: 16519).

\appendix

\section{The anomalous dimension of $\tilde {\cal O}_\pm$}
\label{appA}

In this appendix we compute the anomalous dimension of the operator $\tilde {\cal O}_\pm$ defined in \eqref{bibi1}.
The operator $\tilde {\cal O}_+$
arises, in the large $k$-limit, if we had we considered  a perturbation of the form \eqref{pertparaf} but with a negative relevant sign.
 At the CFT point this should have anomalous
dimension $-\nicefrac2k$ since it corresponds to just the difference of two parafermion bilinears.

\no
Consider the two-point function
\ba
\label{sjksjdskdjks}
&&
\langle \tilde {\cal O}_c(z_1,\z_1) \tilde {\cal O}_c(z_2,\z_2)  \rangle = {1+c^2\ov |z_{12}|^4}
-{\hat g_{12}\ov 2\pi k} \int \text{d}^2z\, \langle \big(\del x_1(z_1) \bar \del x_2(\z_1) + c\, \del x_2(z_1) \bar \del x_1(\z_1)\big)
\nonumber
\\
&&\qq\qquad
\times (\del x_1(z_2) \bar \del x_2(\z_2) + c\, \del x_2(z_2) \bar \del x_1(\z_2))
\big( x_1 x_2 (\del x_1 \bar \del x_2 + \del x_2 \bar \del x_1)\big)(z,\z)\rangle
\nonumber\\
&& \qq\qq\qq
=
{1+c^2\ov |z_{12}|^4}  -{\hat g_{12}\ov 2\pi k} \sum_{i,j,k=1}^2 c^{i+j-2} I_{ijk}  \ ,
\ea
where for convenience we have slightly generalize $\tilde {\cal O}_\pm$ to $\tilde{\cal O}_c(z,\bar z)=\del x_1(z) \bar \del x_2(\z) + c\, \del x_2(z) \bar \del x_1(\z)$ by introducing an arbitrary relative constant $c$.
In the various integrals $ I_{ijk}$ the index structure implies that it arises from the $i$-th, the $j$-th and the $k$-th term
in the first, second and third factors in  the integrand in \eqref{sjksjdskdjks}.  They are given by
\be
\begin{split}
\label{fhj12}
  I_{111} & =\int \text{d}^2z\, \langle \del x_1(z_1)  \del x_1(z_2) \del x_1(z) x_1(z,\z)\rangle
\langle \bar \del x_2(\z_1) \bar \del x_2(\z_2) \bar \del x_2(\z) x_2(z,\z)\rangle
\\
& = \int \text{d}^2z\, \bigg|{1\ov (z-z_1)^2 (z-z_2)} + {1\ov (z-z_2)^2 (z-z_1)}\bigg|^2\,.
\end{split}
\ee
To evaluate the integral we first expand the absolute value
\ba
&& I_{111}=\int\frac{\text{d}^2z}{|z-z_1|^4|z-z_2|^2}+\int\frac{\text{d}^2z}{|z-z_2|^4|z-z_1|^2}
\\
&&\quad +\int\frac{\text{d}^2z}{(z-z_1)^2(z-z_2)(\bar z-\bar z_1)(\bar z-\bar z_2)^2}+
\int\frac{\text{d}^2z}{(z-z_2)^2(z-z_1)(\bar z-\bar z_2)(\bar z-\bar z_1)^2}\,.
\nonumber
\ea
Then we use twice \eqref{second} and \eqref{first} for the first and second line respectively, yielding
\be\label{111}
I_{111}=-\frac{4\pi}{|z_{12}|^4}\left(1+\ln\frac{\varepsilon^2}{|z_{12}|^2}\right)+\frac{4\pi}{|z_{12}|^4}\left(1+\ln\frac{\varepsilon^2}{|z_{12}|^2}\right)=0\,.
\ee
Finally, note that the integrand in \eqref{fhj12} is positive definite. The explanation why the integral turns out to be zero lies in the fact that it is actually divergent and we  may absorb the divergent piece by a field redefinition of the operators $\tilde {\cal O}_\pm$ as explained below 
\eqref{second}. It is easily seen that adding to $\tilde {\cal O}_\pm$ in \eqref{bibi1} 
a term  proportional to ${1\ov\varepsilon^2 k} x_1 x_2$  does the job without giving rise to a mixing with the other operators of 
engineering dimension two, i.e. with ${\cal O}$ and $\tilde {\cal O}$.

In addition, we have that
\be
  I_{112} =\int \text{d}^2z\, \langle \del x_1(z_1)  \del x_1(z_2) \bar \del x_1(\z) x_1(z,\z)\rangle
\langle \bar \del x_2(\z_1) \bar \del x_2(\z_2) \del x_2(z) x_2(z,\z)\rangle \ ,
\ee
has no contribution since it gives rise to contact terms on external points. In addition,
\be
\begin{split}
  I_{211} & =  \int \text{d}^2z\, \langle \bar \del x_1(\z_1)  \del x_1(z_2) \del x_1(z) x_1(z,\z)\rangle
\langle  \del x_2(z_1) \bar \del x_2(\z_2) \bar \del x_2(\z) x_2(z,\z)\rangle
\\
& =  \int {\text{d}^2z \ov |z-z_1|^2 |z-z_2|^4}\,,
\end{split}
\ee
and
\be
\begin{split}
   I_{212} & = \int \text{d}^2z\, \langle \bar \del x_1(\z_1)  \del x_1(z_2) \bar \del x_1(\z) x_1(z,\z)\rangle
\langle  \del x_2(z_1) \bar \del x_2(\z_2) \del x_2(z) x_2(z,\z)\rangle
\\
 & =  \int {\text{d}^2z \ov |z-z_1|^4 |z-z_2|^2}\,,
\end{split}
\ee
Adding the last two integrals and using twice \eqref{second} we find
\be
\label{help.sum}
  I_{211}+ I_{212} =-\frac{4\pi }{|z_{12}|^4}\Big(1+\ln\frac{\varepsilon^2}{|z_{12}|^2}\Big)\,,
\ee
where the divergent pieces have been dismissed (see discussion below \eqref{111}).
Similarly,
\be
\begin{split}
 I_{121} & =\int \text{d}^2z\, \langle \del x_1(z_1) \bar \del x_1(\z_2) \del x_1(z) x_1(z,\z)\rangle
\langle \bar \del x_2(\z_1) \del x_2(z_2) \bar \del x_2(\z) x_2(z,\z)\rangle\\
& =  I_{212}
\end{split}
\ee
and
\be
\begin{split}
I_{122} & =\int \text{d}^2z\, \langle \del x_1(z_1) \bar \del x_1(\z_2) \bar \del x_1(\z) x_1(z,\z)\rangle
\langle \bar \del x_2(\z_1) \del x_2(z_2) \del x_2(z) x_2(z,\z)\rangle\\
& = I_{211}\ .
\end{split}
\ee
Finally,
\be
\begin{split}
I_{221}& =\int \text{d}^2z\, \langle \bar\del x_1(\z_1) \bar \del x_1(\z_2) \del x_1(z) x_1(z,\z)\rangle
\langle  \del x_2( z_1) \del x_2(z_2) \bar \del x_2(\z) x_2(z,\z)\rangle \ ,\\
&=   I_{112}
\end{split}
\ee
and
\be
\begin{split}
 I_{222} & =\int \text{d}^2z\, \langle \bar\del x_1(\z_1) \bar \del x_1(\z_2) \bar \del x_1(\z) x_1(z,\z)\rangle
\langle \del x_2( z_1) \del x_2(z_2) \del x_2(z) x_2(z,\z)\rangle
\\
&  =  I_{111}\ .
\end{split}
\ee
Putting all together in \eqref{sjksjdskdjks} and using \eqref{help.sum}, we find:
\be
\langle \tilde {\cal O}_c(z_1,\z_1) \tilde {\cal O}_c(z_2,\z_2)  \rangle =
{1+c^2\ov |z_{12}|^4}\left(1+\frac{4c}{1+c^2}\frac{\hat g_{12}}{k}\Big(1+\ln\frac{\varepsilon^2}{|z_{12}|^2}\Big)\right)\,.
\ee
Hence, the anomalous dimension of $ \tilde {\cal O}_\pm$ reads
\be
\g^{ \tilde {\cal O}_\pm} =\pm {2\hat g_{12}\ov k}  
= \mp {2 \ov k}{1+\l^2\ov 1-\l^2}\,.
\ee
Hence, the operator $\tilde {\cal O}_+$ has an anomalous dimension 
that matches with \eqref{skfqoqwlql} as expected and now has been explicitly
verified.  The operator $\tilde {\cal O}_-$ corresponds to a marginally irrelevant operator.

\section{Various integrals}
\label{appB}

In this appendix we shall provide the various (single) integrals which appear in the present work. 
Our regularization scheme is as follows: Internal points in integrals can coincide with external ones but coincident external points are not allowed. In addition, to regularize infinities we introduce a very small disc of radius $\varepsilon$ around external points as well.

\no
We shall need the basic integrals
\be
\label{zeroint}
\begin{split}
&\int\frac{\text{d}^2z}{(z-z_1)(\z_2-\z)}=\pi\ln|z_{12}|^2\,,\quad \int\frac{\text{d}^2z}{(z-z_1)(\z_1-\z)}=\pi\ln\varepsilon^2\,,\\
&\int\frac{\text{d}^2z}{(z-z_1)^2(\z_2-\z)}=\frac{\pi}{z_{12}}\,,\quad \int\frac{\text{d}^2z}{(z-z_1)(\z-\z_2)^2}=\frac{\pi}{\z_{12}}\,,
\end{split}
\ee
along with the identity 
\be
\label{partialfractions}
\frac{1}{(z-z_1)(z-z_2)}=\frac{1}{z_{12}}\left(\frac{1}{z-z_1}-\frac{1}{z-z_2}\right)\,.
\ee

\no
Using the above regularization scheme and \eqref{zeroint}, \eqref{partialfractions} we find that
\be
\label{smslsdldsm}
\int\frac{\text{d}^2z}{(z-z_1)(z-z_2)^2(\bar z-\bar z_3)(\bar z-\bar z_4)^2}=-\frac{\pi}{z_{12}^2\bar z_{34}^2}
\left(\ln\frac{|z_{23}|^2|z_{14}|^2}{|z_{13}|^2|z_{24}|^2}+\frac{z_{12}z_{34}}{z_{24}z_{23}}+\frac{\bar z_{12}\bar z_{34}}{\bar z_{13}\bar z_{23}}\right)\,.
\ee
\be
\begin{split}
\label{first}
&\int\frac{\text{d}^2z}{(z-z_1)^2(z-z_2)(\bar z-\bar z_1)(\bar z-\bar z_2)^2}=\frac{2\pi}{|z_{12}|^4}\left(1+\ln\frac{\varepsilon^2}{|z_{12}|^2}\right)\\
&+\frac{\pi}{|z_{12}|^2}\int\text{d}^2z\left(\frac{\z_{12}^{-1}}{(z-z_1)^2(\z-\z_1)}-\frac{z_{12}^{-1}}{(z-z_2)(\z-\z_2)^2}\right)=\frac{2\pi}{|z_{12}|^4}\left(1+\ln\frac{\varepsilon^2}{|z_{12}|^2}\right)\, ,
\end{split}
\ee
since it can be easily seen that in our regularization scheme the integral in the last line of \eqref{first} is zero. For instance, 
$\int{\text{d}^2z\ov  (z-z_1)(\z-\z_1)^2}=\frac{i}{2} \oint_{C_\varepsilon} {\text{d}z\ov |z-z_1|^2}$, where $C_\varepsilon$ is a small circle of
radius $\varepsilon$ surrounding $z=z_1$. Then, the last integral indeed vanishes.
In addition,  using  \eqref{zeroint}, \eqref{partialfractions}
 we find  
\be
\begin{split}
\label{second}
&\int\frac{\text{d}^2z}{|z-z_1|^4|z-z_2|^2}=-\frac{2\pi}{|z_{12}|^4}\left(1+\ln\frac{\varepsilon^2}{|z_{12}|^2}\right)
 + {1\ov |z_{12}|^2} \int{\text{d}^2z \ov |z-z_1|^4} \ .
\end{split}
\ee
Finally, the integral in the last line of \eqref{second} equals $\nicefrac \pi \varepsilon^2$ and is clearly divergent as $\varepsilon\to 0$.  
This integral arises in the computation of the anomalous dimension of the operators  $\tilde {\cal O}_\pm$ 
in appendix \ref{appA}. In order to absorb this divergent piece the operators require a redefinition  similar to that for the single parafermion in  subsection \ref{singlepar}. With this in mind we may safely ignore this term all together and set it to zero.\\
Finally,  using  \eqref{zeroint}, \eqref{partialfractions} we find
\be
\label{third}
\int\frac{\text{d}^2z}{|z-z_1|^2|z-z_2|^2}=-\frac{2\pi}{|z_{12}|^2}\ln\frac{\varepsilon^2}{|z_{12}|^2}\,.
\ee

\end{document}